2/12/2006

# Emergence Explained:

## Getting epiphenomena to do real work

### Russ Abbott

**Department of Computer Science**
**California State University, Los Angeles**
**Los Angeles, California**
**Russ.Abbott@GMail.com**

**Abstract.** Emergence—macro-level effects from micro-level causes—is at the heart of the conflict between reductionism and functionalism. How can there be autonomous higher level laws of nature (the functionalist claim) if everything can be reduced to the fundamental forces of physics (the reductionist position)? We cut through this debate by applying a computer science lens to the way we view nature. We conclude (a) that what functionalism calls the special sciences (sciences other than physics) do indeed study autonomous laws and furthermore that those laws pertain to real higher level entities but (b) that interactions among such higher-level entities is epiphenomenal in that they can always be reduced to primitive physical forces. In other words, epiphenomena, which we will identify with emergent phenomena, do real higher-level work. The proposed perspective provides a framework for understanding many thorny issues including the nature of entities, stigmergy, the evolution of complexity, phase transitions, supervenience, and downward entailment. We also discuss some practical considerations pertaining to systems of systems and the limitations of modeling.

# 1  Introduction

Although the field of complex systems is relatively young, the sense of the term *emergence* that is commonly associated with it—that micro phenomena often give rise to macro phenomena[1]—has been in use for well over a century. The article on *Emergent Properties* in the valuable online Stanford Encyclopedia of Philosophy [O'Connor] begins as follows.

> Emergence [has been] a notorious philosophical term of art [since 1875]. ... We might roughly characterize [its] meaning thus: emergent entities (properties or substances) 'arise' out of more fundamental entities and yet are 'novel' or 'irreducible' with respect to them. ... Each of the quoted terms is slippery in its own right ... . There has been renewed interest in emergence within discussions of the behavior of complex systems.

In a 1998 book-length perspective on his life's work [Holland], John Holland, the inventor of genetic algorithms and one of the founders of the field of complex systems, offered an admirably honest account of the state of our understanding of emergence.

> It is unlikely that a topic as complicated as emergence will submit meekly to a concise definition, and I have no such definition to offer.

In a review of Holland's book, Cosma Shalizi wrote the following.

> Someplace ... where quantum field theory meets general relativity and atoms and

void merge into one another, we may take "the rules of the game" to be given. But the rest of the observable, exploitable order in the universe — benzene molecules, $PV = nRT,$ snowflakes, cyclonic storms, kittens, cats, young love, middle-aged remorse, financial euphoria accompanied with acute gullibility, prevaricating candidates for public office, tapeworms, jet-lag, and unfolding cherry blossoms — where do all these regularities come from? Call this *emergence* if you like. It's a fine-sounding word, and brings to mind southwestern creation myths in an oddly apt way.[2]

The preceding is a poetic echo of the position expressed in a landmark paper [Anderson] by Philip Anderson when he distinguished reductionism from what he called the *constructionist hypothesis,* with which he disagrees, which holds that the

> "ability to reduce everything to simple fundamental laws ... implies the ability to start from those laws and reconstruct the universe"

In a statement that is strikingly consistent with the Stanford Encyclopedia of Philosophy "common understanding" of emergence offered above, Anderson explained his anti-constructionist position.

> At each level of complexity entirely new properties appear. ... [O]ne may array the sciences roughly linearly in [the following] hierarchy [in which] the elementary entities of [the science at level n+1] obey the laws of [the science at level n]: elementary par-

---

[1]  Recently the term *multiscale* has gained favor as a less mysterious-sounding way to refer to this macro-micro interplay. The fact that *multiscale* sounds more scientific, however, does not reflect a correspondingly clearer understanding of the phenomenon itself.

[2]  Shalizi also offers his own definition of emergence on his website [Shalizi] as follows.

One set of variables, *A*, emerges from another, *B* if (1) *A* is a function of *B*, i.e., at a higher level of abstraction, and (2) the higher-level variables can be predicted more efficiently than the lower-level ones, where "efficiency of prediction" is defined using information theory.

ticle physics, solid state (or many body) physics, chemistry, molecular biology, cell biology, ..., psychology, social sciences. But this hierarchy does not imply that science [n+1] is 'just applied [science n].' At each [level] entirely new laws, concepts, and generalization are necessary. ... Psychology is not applied biology, nor is biology applied chemistry. ... The whole becomes not only more than but very different from the sum of its parts.

Although not so labeled, the preceding provides a good summary of the position known as functionalism (or in other contexts as non-reductive physicalism), which argues that autonomous laws of nature appear at many levels.

Anderson thought that the position he was taking was radical enough—how can one be a reductionist, which he claimed to be, and at the same time argue that there are autonomous sciences—that it was important to reaffirm his adherence to reductionism. He summed this up as follows.

"[The] workings of all the animate and inanimate matter of which we have any detailed knowledge are all ... controlled by the same set of fundamental laws [of physics]. ... [W]e must all start with reductionism, which I fully accept."

In the rest of this paper, we elaborate and extend the position that Anderson set forth. We hope to offer a coherent explanation for how nature can be both reductive and non-reductive simultaneously.

Much of our approach is derived from concepts borrowed from Computer Science—which more than any other human endeavor has had to deal, on the most concrete terms, with the question of how one can operate on many levels simultaneously. How is it possible for such the amazing panoply of functionality created by software to emerge from (nothing but) electrons in motion?

The rest of this paper is organized as follows.

- Section 2 develops basic concepts. It explores the notions of reductionism and functionalism, and it characterizes their differences and points of agreement. It defines the term *epiphenomenon*. It explicates the notion of supervenience and points out an important limitation. It argues that one must chose between force reductionism and the position that new forces of nature come into being mysteriously.

- Section 3 uses the Game of Life to illustrate and then to define emergence.

- Section 4 explores some of the implications of our definition. It defines the notion of downward entailment. It discusses the reality of higher level abstractions, and it offers a novel view of phase transitions.

- Section 5 defines the notion of an entity as a persistent region of reduced entropy. It relates the concepts of entities, dissipative structures, and autonomy. It shows why emergence is a fundamental feature of nature. It distinguishes natural from artificial autonomous entities. It shows why supervenience is not as powerful a concept as one might have hoped. It discusses the conceptual limitations Computer Science suffers as a result of its self-imposed exile to a world of free energy.

- Section 6 discusses stigmergy, historical contingency, and the evolution of complexity.

- Section 7 presents additional implications for science of entities, emergence, and complexity.

- Section 8 presents a framework for the varieties of emergence that we discuss.

- Section 9 offers some practical advice. about service-oriented architectures, stove-piped systems, and the limitations of modeling.

- Section 10 provides brief a summary and includes a remark about an area for future investigation.

- The Appendix offers a formal definition of Games of Life patterns such as the glider. It shows how such patterns can be used to create an API of patterns. It presents some basic unsolvability results.

## 2   Background and foundations

We begin by contrasting reductionism and functionalism. We use papers written by Steven Weinberg, a reductionist physicist, and Jerrold (Jerry) Fodor, a functionalist philosopher, as our points of departure.

- Weinberg, a professor of physics at the University of Texas, Austin, was awarded the Nobel prize in physics for his work on unifying the weak and electromagnetic forces. He is an articulate spokesperson for a constructivist-like form of reductionism, the position that all of science can be reduced to and are mathematical consequences of the laws of physics. He presented his views in "Reductionism Redux" [Weinberg], an essay in which he responded to position papers of participants in a debate on reductionism held in 1992 at Jesus College, Cambridge University.

- Fodor, a professor of philosophy at Rutgers, is one of the founders of functionalism, the position that regularities appear at all levels of science and that these regularities are not reducible to physics. He reviewed [Fodor 98] his position in "Special Sciences; Still Autonomous after All These Years," which he wrote in reply to a more reductionist position taken [Kim '92, '93] by Jaegwon Kim, a professor of philosophy at Brown University.

In the end we will claim that it was neither the reductionist Weinberg nor the functionalist Fodor but Anderson—who proposed a marriage between reductionism and functionalism—who was right.[3] Our job will be to show how the improbable couple can live happily ever after.

### 2.1   Functionalism

Functionalism [Fodor 74] holds that there are so-called 'special sciences' (in fact, all sciences other than physics and perhaps chemistry) that study regularities in nature that are in some sense autonomous of physics. In [Fodor 98] Fodor wrote the following reaffirmation of functionalism.

> The *very existence* of the special sciences testifies to the reliable macrolevel regularities that are realized by mechanisms whose physical substance is quite typically heterogeneous. Does anybody really doubt that mountains are made of all sorts of stuff? Does anybody really think that, since they are, generalization about mountains-as-such won't continue to serve geology in good stead? Damn near everything we know about the world suggests that uni-maginably complicated to-ings and fro-ings of bits and pieces at the extreme *micro*level manage somehow to converge on stable *macrolevel* properties.

---

[3]   All of the extracts from Anderson, Fodor, and Weinberg are from the three papers cited above. The emphases in the extracts are all in the originals.

Although Fodor does not use the term, the phenomena studied by the special sciences are the same sort of phenomena that we now call multiscale, i.e., emergent.

Why is there emergence? Fodor continues as follows.

> [T]he 'somehow' [of the preceding extract] really is entirely mysterious … .
>
> So, then, *why is there anything except physics?* … Well, I admit that I don't know why. I don't even know how to think about why. I expect to figure out why there is anything except physics the day before I figure out why there is anything at all … .

So, like Holland, Fodor throws up his hands with respect to explaining emergence.

One of the tenets of functionalism is that within any domain, it is the regularities that appear at the level of phenomena with which the domain is concerned that are important. It is not significant how those regularities are realized in terms of lower level phenomena—both because that doesn't matter and because they can often be implemented in any of a number of ways. A term commonly used in Functionalism is *multiple realizability*, which refers to the notion that many regularities (many functions) can be realized in multiple ways. As Fodor remarks,

> that's why references to can openers, mousetraps, camshafts, calculators and the like bestrew the pages of functionalist philosophy. To make a better mousetrap *is* to devise a new kind of mechanism whose behavior is reliable with respect to the

high-level regularity "live mouse in, dead mouse out."

## 2.2 Reductionism

Taking the other side of the debate is Steven Weinberg, one of the most articulate defenders of reductionism. Weinberg distinguishes two kinds of reductionism.

> We ought first of all to distinguish between what (to borrow the language of criminal law) I like to call grand and petty reductionism. Grand reductionism is … the view that all of nature is the way it is (with certain qualifications about initial conditions and historical accidents) because of simple universal laws, to which all other scientific laws may in some sense be reduced. Petty reductionism is the much less interesting doctrine that things behave the way they do because of the properties of their constituents: for instance, a diamond is hard because the carbon atoms of which it is composed can fit together neatly. …
>
> Petty reductionism is not worth a fierce defense. … In fact, petty reductionism in physics has probably run its course. Just as it doesn't make sense to talk about the hardness or temperature or intelligence of individual "elementary" particles, it is also not possible to give a precise meaning to statements about particles being composed of other particles. We do speak loosely of a proton as being composed of three quarks, but if you look very closely at a quark you will find it surrounded with a cloud of quarks and anti-quarks and other particles, occasionally bound into protons; so at least for a brief moment we could say that the quark is made of protons.

Weinberg continues his explication of grand reductionism by using the weather as an example.

> [T]he reductionist regards the general theories governing air and water and radiation as being at a deeper level than theories about cold fronts or thunderstorms, not in the sense that they are more useful, but only in the sense that the latter can in principle be understood as mathematical consequences of the former. The reductionist program of physics is the search for the common source of all explanations. ...

Reductionism ... provides the necessary insight that there are no autonomous laws of weather that are logically independent of the principles of physics. ... We don't know the final laws of nature, but we know that they are not expressed in terms of cold fronts or thunderstorms. ...

Every field of science operates by formulating and testing generalizations that are sometimes dignified by being called principles or laws. The library of the University of Texas has thirty-five books with the title "Principles of Chemistry" and eighteen books with the title "Principles of Psychology". But there are no principles of chemistry that simply stand on their own, without needing to be explained reductively from the properties of electrons and atomic nuclei, and in the same way there are no principles of psychology that are freestanding, in the sense that they do not need ultimately to be understood through the study of the human brain, which in turn must ultimately be understood on the basis of physics and chemistry.

Thus the battle is joined: are the higher level sciences derived from physics?

Before approaching this question, it is reasonable to ask whether any common ground exists between reductionism and functionalism? To do that, we first examine the concepts of epiphenomena and supervenience. Then we explore the status of the fundamental forces of physics and the possibility of the emergence of higher level forces. Finally we look at the role the environment and historical accidents play in both reductionism and functionalism.

## 2.3 Epiphenomena

If one doesn't already have a sense of what it means, the term *epiphenomenon* is quite difficult to understand. Here is the WordNet definition [WordNet], which is representative.

> A secondary phenomenon that is a by-product of another phenomenon.

It is not clear that this definition pins much down. This definition is especially troublesome because the terms *secondary* and *by-product* should not be interpreted to mean that an epiphenomenon is separate from and a consequence of the state of affairs characterized by the "other" phenomenon.

We suggest that a better way to think of an epiphenomenon is as an alternative way of apprehending or perceiving a given state of affairs. Consider Brownian motion, which appears to be motion that very small particles of non-organic materials are able to engage in on their own. Before Einstein, Brownian motion was a mystery. How could inanimate matter move on its own? We now know that Brownian motion is *an epiphenomenon of* collisions of particles with atoms or molecules.

With this usage of *epiphenomenon* as a guide we define an epiphenomenon as a phenomenon that can be described (sometimes formally but sometimes only informally) in terms that do not depend on the underlying phenomena from which it emerges.[4]

---

[4] We use the term *emerges* advisedly. We will soon define *emergent* and *epiphenomenal* to be synonyms.

This is familiar territory for Engineering and Computer Science. Requirements and specifications are by intention written in terms that do not depend on the design or implementation of the systems that realize them. Requirements are written before systems are designed, and specifications are intended specifically to be implementation-independent.

Although use of the term *epiphenomena* in this context sounds strange, a requirements document or a system specification is intended to describe epiphenomena of systems that satisfy those requirements or that meet that specification.[5]

Even though we just claimed that a specification describes epiphenomena, our notion of epiphenomena is not the same as the notion of functionality. From our perspective epiphenomena exist only when there is an implementation. Epiphenomena are, first of all, phenomena. A specification for a system that did not exist would not describe epiphenomena. It would describe the epiphenomena of any system that satisfies that specification, but that's as far as one could go. From our perspective, there are no epiphenomena unless they are epiphenomena of something. That is not the case with functionality, which is understood in the abstract.

It is reasonable to say that the functionality of most executing software is epiphenomenal. The computation considered as an abstraction may be defined independently of the implementation. But the only real action is at the very lowest level. No matter how abstract one's software, one can always stop a computation by pulling the plug on the computer. The computation as an abstract epiphenomenon exists only because electrons are actually flowing.

## 2.4 Supervenience

A closely related term from the philosophical literature is *supervenience*. The intended use of this term is to relate a presumably higher level set of predicates or properties[6] (call this set H for *higher*) to a presumably lower level set of predicates or properties (call this set L for *lower*). The properties or predicates in H and L are all presumed to be applicable to some common domain of discourse.

H and L are each ways of characterizing the state of affairs of the underlying domain. For any particular state of affairs in the domain of discourse, the predicates in H and L will each be either true or false (or perhaps not applicable).

One says that H supervenes on (or over) L if it is never[7] the case that two states of affairs will assign the same configuration of values to the elements of L but different configuration of values to the elements of H. In other words when a state of affairs assigns values to predicates in L, that fixes the assignments of values to predicates of H.

Consider the following simple example. Let the domain be a sequence of n bits. Let L be the statements: bit 1 is on; bit 2 is on; etc. Let H be statements of the sort: exactly 5 bits are on; an even number of bits are on; no two successive bits are on; the bits that are on form the initial values in the Fibonacci sequence; etc.

---

[5]  We are not holding our breath waiting for a contracting officer to refer to a system's inadequate epiphenomena.

[6]  Since properties (e.g., the color of an object) may be expressed in terms of predicates about that property, from here on we will speak only of predicates.

[7]  Some definitions require that not only is it never the case, it never can be the case. It does make a formal difference whether we base supervenience on a logical impossibility or on empirical facts. We finesse that distinction by adopting the rule of thumb of fundamental particle physicists: if something can happen it will.

H supervenes on L since any configuration of values of the statements in L determines the values of the statements in H.

However, if we remove one of the statements from L, e.g., we don't include in L a statement about bit 3, but we leave the statements in H alone, then H does not supervene on L.

To see why, consider the H statement

$h_1$: an even number of bits is on.

For concreteness, let's assume that there are exactly 5 bits. Let's assume first, as in the first line of Figure 1, that all the bits except bit 3, the one for which there is no L statement, are on. Thus since there is no L statement about bit 3, all the L statements are true even though bit 3 is off. Since 4 of the 5 bits are on, $h_1$ is also true.

Now, assume that bit 3 is on as in the second line of Figure 1. All the L statements are still true. But since 5 bits are now on, $h_1$ is now false.

Since we have found an H statement that has two different values for a single configuration of values of the L statements, H does not supervene over L.

Although we have not attempted to trace its history, the notion of supervenience may have originated as an attempt to capture the relationship between epiphenomena and their underlying phenomena. Presumably epiphenomena supervene on underlying phenomena: distinct epiphenomena must be associated with distinct underlying phenomena, which is

what one wants.[8] You can't get two different sets of epiphenomena from the same underlying phenomena.[9]

Note that the reverse is not true. Two different states of the underlying phenomena may result in the same epiphenomena. In our bit example, there are many different ways in which an even number of bits may be on.

The position known as supervenience physicalism might be understood as claiming that any higher-level description of nature supervenes over some set of primitive descriptions.

It would appear that the relationship defined by supervenience will be useful in analyzing multi-scale phenomena. To some extent this is the case. When we consider autonomous entities, however, we shall see that supervenience is not as useful as one might have hoped. This is one reason that emergence has been so difficult to pin down.

One reason that supervenience is less useful than one might have hoped may be related to the difficulty one encounters when using supervenience for infinite domains. Consider our bit example again, but imagine that we have a countably infinite number of bits. Consider the H statement

---

[8]   On the other hand, searches for the phrase "epiphenomena supervene" on Google, Yahoo msn.com, and AskJeeves conducted on 9/17/2005 found no references—other than to an online draft of this paper. This suggests that none of these services had ever scanned a document that contained the sentence fragment "… epiphenomena supervene over … ."

[9]   It's not really the epiphenomena that supervene over their underlying phenomena. It's statements about the epiphenomena or properties of the epiphenomena that supervene over statements about the underlying phenomena or their properties. As a short-hand, however, we will talk about epiphenomena as supervening (or not) over underlying phenomena.

$h_2$: the bits that are on are prime.

Clearly the H set consisting solely of $h_2$ supervenes over the entire set of L statements. Just as clearly, that H set does not supervene over any proper subset of the L statements, and certainly not over any finite subset of the L statements—one needs to look at all of the bits to determine whether it is exactly the prime bits that are on.

So even though we can conclude that H, which contains a single relatively simple statement, supervenes over the infinite set of statements in L, that information doesn't buy us much. On the contrary, supervenience of this sort is like describing a tapestry by enumerating the threads that make it up. The epiphenomena are lost. We will see another example of this later.

## 2.5 Fundamental forces and strong emergence

Returning to Weinberg and Fodor, presumably both would agree that phenomena of the special sciences supervene on phenomena in physics. A given set of phenomena at the level of fundamental physics is associated with no more than one set of phenomena at the level of any of the special sciences. Or looking top-down, two different states of affairs in some special science must be associated with two different states of affairs at the level of fundamental physics. Supervenience of this sort seems to correspond more or less to Weinberg's petty reductionism, a doctrine that he finds of only minor importance. So perhaps agreement at this level is not very significant.

Where Weinberg and Fodor disagree is not about supervenience but about whether the principles of the special sciences can be derived from the principles of physics.

That disagreement aside, we do have a fundamental area of agreement—which has some quite significant implications. Weinberg makes his case sarcastically.

> Henry Bergson and Darth Vader notwithstanding, there is no life force. This is [the] invaluable negative perspective that is provided by reductionism.

What I believe Weinberg is getting at is that the current standard model of physics postulates four elementary forces: the strong force, the weak force, the electromagnetic force, and gravity. Since Weinberg's Nobel prize was for his work on unifying the weak and electromagnetic force, perhaps one should say there are only three fundamental forces. Either way, what's important is that according to physics there is a small fixed number of fundamental forces. I doubt that Fodor would disagree.

Weinberg's sarcastic reference to a life force is an implicit criticism of an obsolete strain of thinking about emergence. The notion of vitalism—the emergence of life from lifeless chemicals—postulates a new force of nature that appears at the level of biology and that is not reducible to lower level phenomena. Emergence of this sort is what Bedau [Bedau] has labeled "strong emergence." As Bedau also points out, no one takes this kind of emergence seriously.

It is worth noting, however, that even were evidence of strong emergence to be found, science would not shrivel up and die. Dark energy, the apparently extra force that seems to be pushing the Universe to expand may be a new force of nature.

Furthermore, even if other (spooky) forces of nature like vitalism were (mysteriously) to appear at various levels of complexity, science would carry on. We would do our best to understand such forces by measuring and characterizing them in any way that we could.

What one doesn't want is to have new forces of nature popping up indiscriminately. Strong emergence would not be so terrible as long as it didn't happen too often—after all, the existing primitive forces just seemed to pop up out of nowhere and we have taken them in stride—or if it did happen often, if we could find a theory that let us predict when they would appear.

What would be especially upsetting would be a new force of nature that violated the conservation laws. Suppose one could put certain components together and produce energy that didn't depend on the energy or mass of the components—the mass and energy of the components remained intact, but the combination of those components created new energy. That would be scientifically quite disturbing—although it would certainly help with the energy problem.

Since the appearance of new forces of nature that do not upset the conservation laws could be taken simply as an acceptable extension of physics (but one that we have not seen and do not know how to produce), and since the appearance of new forces that violate the conservation laws is currently unimaginable, from now on we ignore the possibility of strong emergence.

If one dismisses the possibility of strong emergence and agrees that the only forces of nature are the fundamental forces as determined by physics, then Fodor must also agree (no doubt he would) that any force-like construct postulated by any of the special sciences must be strictly reducible to the fundamental forces of physics. As Weinberg says, there is no life force.

Note that this is a truly stark choice: strict reductionism with respect to forces or strong emergence. There is no third way.

This leads to an important conclusion. Any cause-like effect that results from a force-like phenomenon in the domain of any of the special (i.e., higher level) sciences must be epiphenomenal.[10] Since epiphenomenal forces supervene on fundamental forces, distinct epiphenomenal interactions must be manifestations of distinct fundamental force actions. In other words, anything that happens at the most fundamental level (if there is one) has no more than one manifestation at any higher level.

It is important to note that this perspective establishes one of the basic claims of reductionism: forces at all levels must be explicable in terms of—i.e., they are epiphenomenal of and reducible to—the fundamental forces of physics.[11] There are no magical mystery forces; there is no strong emergence.

The fact that all special (i.e., higher level) science interactions are epiphenomenal is not a problem for functionalism. Nor is it news to Fodor, who speaks freely of "the to-ings and fro-ings of bits and pieces at the extreme *micro*level." Rather, the functionalist claim is that the regularities (be they epiphenomenal or not) that appear at the level of any special science are of significance on their own. The fact that the interactions to which one refers in describing those regularities are in fact implemented by lower level interactions does not diminish either the importance or the lawfulness of the higher level regularities—even though as we have seen Fodor has no idea how those regularities come about.

---

[10] Kim [Kim '93] (denigratingly) used the term *epiphenomenal causation* to refer to interactions of this sort. We also consider such interactions to be epiphenomenal, but we don't find them less worthy as a result.

[11] Compare this with the conclusion Hume reached [Hume] in his considerations of causality—that when one looks carefully at any allegedly direct causal connection, one will find intermediary links. Since Hume did not presume what we now consider to be a bottom level of fundamental physical forces, he dismissed the notion of causality entirely.

### 2.6 Historical accidents and the environment

There is a second area of at least implicit agreement between Weinberg and Fodor. Consider the following from Weinberg.

> [A]part from historical accidents that by definition cannot be explained, the [human] nervous system [has] evolved to what [it is] entirely because of the principles of macroscopic physics and chemistry, which in turn are what they are entirely because of principles of standard model of elementary particles.

Note Weinberg's reference to historical accidents—which we also saw earlier, in both his definition of grand reductionism and his discussion of the weather. Weinberg gives historical accidents as important a role in shaping the world as he gives to the principles of physics. But he gives them a significantly lesser billing—one might say below the title instead of above.

The importance of historical accidents is especially clear when thinking about evolution. We suggest that contrary to Weinberg's claim, the human nervous system (and human anatomy in general) evolved to what they are not primarily because of the principles of macroscopic physics and chemistry but primarily because of the environment in which that evolution took place—and in which the nervous system and anatomy must function.

Because of the crucial role the environment plays in evolution, one cannot start with the principles of physics and chemistry and, considering them in isolation, derive human anatomy. Certainly human anatomy must be consistent with the principles of physics and chemistry. But human anatomy cannot be derived exclusively from the principles of physics and chemistry. One must also consider the environment in which that anatomy is intended to function.

We don't think we are putting words into Weinberg's mouth when we take his references to "historical accidents" to mean the environment and the context in which something occurs. If one grants us this, then as Weinberg acknowledges, the environment is often as significant a consideration in how things turn out as are the principles of physics.

The same perspective is nearly (but not quite) explicit in the extracts from Fodor. Recall Fodor's list of artifacts, including a mouse trap, a can opener, etc. All those artifacts are defined functionally, i.e., in terms of the functions they perform—which necessarily means functions performed in the environment in which they exist.

A can opener in an environment in which cans are of an entirely different size or shape from our cans would not be a can opener. The same is true for a mousetrap in an environment in which mice were either the size of fleas or the size of elephants.

And of course, the very name of the school of thought, functionalism, underlines its concern with how things function. Functionality is by definition a rela-tionship between something and its environment—even if we limit the notion of an environment to the "inputs" and "outputs" of the element under consideration.

It is also important to note that when one speaks of an environment, one must understand the environment at the level at which the object under consideration interacts with it—even though as we saw earlier those interactions are epiphenomenal. A can opener opens cans, and a mouse trap traps mice. Neither deals with swarms of quarks and other fundamental particles. As Fodor says, "live mouse in, dead mouse out."

Although neither side of this debate focuses on this issue, they are both apparently in agreement that the environment within which something exists is important.

Interaction with an environment will turn out to be quite significant. We explore it further when we discuss autonomous entities and stigmergy.

## 3  Emergence in the Game of Life

In this section we use the Game of Life[12] [Gardner] to illustrate emergence: the implementation of a new level of abstraction on top of an existing substrate.

- The Game of Life is a totalistic[13] two-dimensional cellular automaton.

---

[12]  The Game of Life is a popular example in discussions of emergence. Bedau uses it as the primary example in "Downward causation and the autonomy of weak emergence" [Bedau]. We return to Bedau later. Dennett refers to it in "Real Patterns" [Dennett] when discussing how his intentional stance perspective compares to the perspectives of other philosophers with respect to the reality of beliefs. (His position is that beliefs should be considered to be "mildly real," which is intermediate within the spectrum of positions he examines.)

[13]  *Totalistic* means that the action taken by an agent depends on the number of its neighbors in certain

The Game of Life grid is assumed to be unbounded in each direction, like the tape of a Turing Machine.

- An agent occupies each grid cell. Agents are fixed and cannot move around on the grid. This is not unusual in agent-based modeling.

- Each agent is in one of two states: "alive" or "dead" or more simply on or off.

- The 8 surrounding agents are an agent's neighbors.

- At each time step an agent determines whether it will be alive or dead at the next time step according to the following rules.

  - A live agent with two or three live neighbors stays alive; otherwise it dies.

  - A dead agent with exactly three live neighbors is (miraculously) (re)born and becomes alive.

- All agents update themselves simultaneously based on the values of their neighbors at that time step.

It is useful to think of the Game of Life in the following three ways.

1. Treat the Game of Life is a serious agent-based model—of something, perhaps life and death phenomena. For our purposes it doesn't matter that the Game of Life isn't a realistic model—of anything. Many agent-based models are at the same time quite simple and quite revealing.

2. Treat the Game of Life as a trivial physical universe. Recall Shalizi:

---

states and not on which neighbors are in which state.

Someplace … where quantum field theory meets general relativity … we may take "the rules of the game" to be given."

The Game of Life rules will be those "rules of the game." The rules that determine how cells turn on and off will be taken as the most primitive operations of the physics of the Game of Life universe.[14] When thinking in terms of this perspective, rather than thinking of a grid cell as occupied by (immobile) agents, we think of the cells themselves as something like primitive particles.

The reductionist agenda within such a Game of Life universe would be to reduce every higher level phenomenon to the underlying Game of Life rules.

3. Treat the Game of Life as a programming platform.

Although these three perspectives will yield three different approaches to the phenomena generated, the phenomena themselves will be identical. It will always be the same Game of Life rules which determine what happens.

### 3.1  Epiphenomenal gliders

Figure 2 shows a sequence of 5 time steps in a Game of Life run. The dark cells (agents) are "alive;" the light cells (agents) are "dead." One can apply the rules manually and satisfy oneself that they produce the sequence as shown.

Notice that the fifth configuration shows the same pattern of live and dead cells as the first except that the pattern is offset

---

[14] This is the basis of what is sometimes called "digital physics" (see [Zuse], [Fredkin], and [Wolfram]), which attempts to understand nature in terms of cellular automata.

---

by one cell to the right and one cell down.

If there are no other live cells on the grid, this process could be repeated indefinitely, producing a glider-like effect.

Such a glider is an epiphenomenon of the Game of Life rules. If one thinks about it—and forgets that one already knows that the Game of Life can produce gliders—gliders are quite amazing. A pattern that traverses the grid arises from very simple (and local) rules for turning cells on and off.

We should be clear that gliders are epiphenomenal. The rules of the Game of Life do nothing but turn individual cells on and off. There is nothing in the rules about waves of cells turning on and off sweeping across the grid. Such epiphenomenal gliders exemplify emergence.

- Gliders are not generated explicitly: there is no glider algorithm. There is no "code" that explicitly decides which cells should be turned on and off to produce a glider.

- Gliders are not visible in the rules. None of the rules are formulated, either explicitly or implicitly, in terms of gliders.

When looked at from our agent-based modeling perspective, gliders may represent epidemics or waves of births and deaths. If one were attempting to demonstrate that such waves could be generated by simple agent-agent interactions, one might be quite pleased by this result. It might merit a conference paper.

### 3.2  Gliders in our physics world

From our physics perspective, we note that the rules are the only forces in our Game of Life universe. Being epiphenomenal, gliders are causally power-less.[15] The existence of a glider does not change either how the rules operate or which cells will be switched on and off.

Gliders may be emergent, but they do not represent a new force of nature in the Game of Life universe. It may appear to us as observers that a glider looks like it is moving across the grid and that when it reaches a certain cell it will turn that cell on.

But that's not true. It is only the rules that turn cells on and off. A glider doesn't "go to an cell and turn it on." A Game of Life run will proceed in exactly the same way whether one notices the gliders or not. This is a very reductionist position. Things happen only as a result of the lowest level forces of nature, which in this case are the rules.

### 3.3 The Game of Life as a programming platform

Amazing as they are, gliders are also trivial. Once one knows how to produce a glider, it's a simple matter to make as many as one wants. If we look at the Game of Life as a programming platform—imagine that we are kids fooling around with a new toy—we might experiment with it to see whether we can make other sorts of patterns. If we find some, which we will, we might want to see what happens when patterns crash into each other—boys will be boys.

After some time and effort, we might compile a library of Game of Life patterns, including the API[16] of each pattern,

which describes what happens when that pattern collides with other patterns.[17]

Since its introduction three decades ago, a community of such Game of Life programmers has developed. That community has created such libraries—at least on an informal basis.[18]

It has even been shown [Rendell] that by suitably arranging Game of Life patterns, one can simulate a Turing Machine.

Moreover, and this is a crucial point, the emulation of a Turing Machine with Game of Life patterns is also an example of emergence. There is no algorithm. The Turing Machine appears as a consequence of epiphenomenal interactions among epiphenomenal patterns!

What did we just say? What does it mean to say that epiphenomenal gliders and other epiphenomenal patterns simulate a Turing Machine? How can it mean anything? The patterns aren't real; the Turing Machine isn't real; they are all (nothing but) epiphenomena.

Furthermore, even the interactions between and among patterns aren't real either. They're also epiphenomenal—and epiphenomenal in the sense described above: the only real action is at the most fundamental level, the Game of Life rules. Pattern APIs notwithstanding the only thing that happens on a Game of Life grid is that the Game of Life rules determine which cells are to be on and which cells are to be off. No matter how real the patterns look to us, interaction among them is always epiphenomenal. So what are we talking about?

---

[15]  All epiphenomena are causally powerless. Since epiphenomena are simply another way of perceiving underlying phenomena, an epiphenomenon itself cannot have an effect on anything. It is the underlying phenomena that act.

[16]  Application Programming Interface

---

[17]  Note, however, that interactions among patterns are quite fragile. If two patterns meet in slightly different ways, the results will generally be quite different.

[18]  Many of these libraries are available on the web. To explore this world, a good place to start and one that seems to be kept up to date is *Jason's Life Page* [Summers]. The patterns available on that page and on the pages to which that page links are quite amazing.

What does one do to show that a Game of Life emulation of a Turing machine is correct? What one must do is to adopt a operational perspective and treat the patterns and their interactions, i.e., the design itself, independently of the Game of Life.

It is the design, i.e., the way in which the patterns—be they real or epiphenomenal—interact that we want to claim simulates a Turing Machine. To show that we must do two things.

1. Show that the abstract design consisting of patterns and their interactions actually does simulate a Turing Machine. That is, we reify the design, i.e., treat it as real, and argue about its properties.

2. Show that the design can be implemented on a Game-of-Life platform.

Note what this perspective does. It unshackles the design from its moorings as a Game of Life epiphenomenon, reifies it as an independent abstraction, and lets it float free. (The protestors in the streets chanting "Free the design" can now lower their picket signs and go home.)

The design becomes an abstraction, an abstract construct on its own. Once we have such a design as an abstraction we can then reason about its properties, i.e., (a) that it accomplishes what we want, namely that it simulates a Turing Machine and (b) that it can be reattached to its moorings and be implemented on a Game of Life platform. In other words, emergence is getting epiphenomena to do real (functional) work.

### 3.4  Game of Life anthropologists

Let's forget everything we just said about the Game of Life, and let's pretend we are anthropologists. Let's imagine that a lost tribe of what turn out to be Game of Life creatures has been discovered in a remote wilderness. Of course, when they are first discovered, we don't know that they are Game of Life creatures. All we know is that their strange grid-like faces are made up of cells that blink on and off.

We get a grant to study these creatures. We travel to their far-off village, and we learn their language. They can't seem to explain what makes their cells blink on and off; we have to figure that out for ourselves.

After months of study, we come up with the Game of Life rules as an explanation for how the grid cells are controlled. That seems to work. Every single member of the tribe operates in a way that is consistent with those rules. The rules even explain the unusual patterns that we observe—some of them, glider-like, traversing the entire grid. Thrilled with our analysis, we return home and publish our results.

But one thing continues to nag. One of the teenage girls—she calls herself Hacka for reasons that we do not understand—has a pattern of activities on her grid that seems more complex than most of the others. The Game of Life rules fully explain every light that goes on and every light that goes off on Hacka's pretty face. But somehow that explanation doesn't seem to capture everything that's going on. It just seems more complex than that. Did we miss something?

To make a long story short, it turns out that the tribe was not as isolated as we had thought. In fact they have an Internet connection. Hacka had learned not only that she was a Game of Life system but that the Game of Life can emulate a Turing Machine. She had decided to program herself to do just that. Her parents disapproved. But girls just want to have fun—especially teenage girls.

No wonder we felt uncertain about our results. Even though the Game of Life rules explained every light that went on and off on Hacka's face, it said nothing about the functionality implemented by Hacka's Turing Machine emulation.

The rules explained everything about how the system worked; but they said nothing about what the system did. The rules didn't have a way even to begin to talk about the functionality of the system—which was logically independent of the rules. The rules simply don't talk about Turing Machines.

A Turing machine is an autonomous functional abstraction that we (and Hacka) built on top of the rules of the Game of Life. Our reductive explanation, that a certain set of rules make the cells go on and off, had no way to capture this sort of additional functionality.[19]

### 3.5   Keeping score

In the debate between reductionism and functionalism, the fact that one can build a Game of Life Turing Machine may be scored as follows.

- Reductionism scores a point—with which it was credited earlier—in that the only forces operating are (and must be) reducible to the fundamental operational rules of the Game of Life.

  In philosophical terms, this would be considered an supervenience physicalism: any apparent higher level forces or interactions are nothing more than conceptually convenient ways of packaging lower level forces and interactions.

- Functionalism scores a point in that the functionality of a Turing Machine is beyond the realm of and can neither be reduced to nor derived from the Game of Life rules.

  Functionalism gets only partial credit because it is quite clear (if immensely complicated)—but not at all mysterious and not on a par with the question of why anything exists at all—how a Turing Machine may be implemented in  terms of Game of Life rules. All that is required is that a design for a Turing Machine emulator be implemented by using the Game of Life as a software development platform.

### 3.6   Defining emergence

With this background, we can define emergence as a relationship between a phenomenon and a model.

By a model we will mean a collection of elements with certain interrelationships, e.g., the grid cells and rules of the Game of Life.

To begin, given some model we will say that a phenomenon is emergent over that model if it is epiphenomenal with respect to that model. In other words, all epiphenomena are emergent, and all

---

[19]  In "Real Patterns" Dennett [Dennett '91] uses the fact that a Turing Machine may be implemented in terms of Game of Life patterns to argue that the position he takes in *The Intentional Stance* [Dennett '87] falls midway along a spectrum of positions ranging from what he calls "Industrial strength Realism" to eliminative materialism (that beliefs are nothing but convenient fictions).

Our focus in this paper differs from Dennett's in that it is not on psychological states or mental events but on the nature of regularities—independently of whether those regularities are the subject matter of anyone's beliefs.

emergent phenomena are epiphenome-nal. In short, we define *epiphenomenal* and *emergent* to be synonyms.

Recall that we defined an epiphenome-non to be a phenomenon that is appre-hended, conceptualized, or perceived in-dependently of the forces that bring it about. Thus a phenomenon is emergent if it is conceptualized independently of the phenomena that implement it. As we pointed out in our discussion of epiphe-nomena earlier, this is familiar territory for Engineering and Computer Science. Requirements and specifications are by intention written in terms that do not de-pend on the design or implementation of the systems that realize them.

More concretely we define a phenome-non as emergent over a model if it satis-fies two conditions.

1. The phenomenon as a phenomenon may be understood on its own. Its conceptualization does not depend on the model from which it emer-ges.[20]

2. There is an implementation of the phenomenon in terms of elements of the model.[21]

Our prototypical examples are gliders and Turing Machines, which we will say are emergent over the Game of Life.

---

[20] We are not requiring that there be a person to do the conceptualizing, only that such an independent conceptualization be possible.

[21] Our concern with implementations distinguishes our position from both functionalism and Den-nett's intentional stance, neither of which cares much about how higher level phenomena are im-plemented.

It may be that emergent phenomena may have multiple realizations—Turing Machines may be epiphenomena of many computing substrates—but this will not be the most commonly occurring situation in nature. Non-computational epiphe-nomena will usually occur as the consequence of a particular implementation.

Notice that we do not require an emer-gent phenomenon to have a formal de-scription or specification. Certainly some epiphenomena may be formalized and understood as formalized abstrac-tions—the Turning Machine for exam-ple. Epiphenomena about which such in-dependent formal theories may be de-veloped often tend to be particularly use-ful cases of emergence from a scientific or mathematical perspective.

On the other hand, we shall see later that there are epiphenomena about which it is either not feasible or not useful to define abstract theories but which are important to us nevertheless. As an immediate and familiar (if informal) example, consider the so-called edge-of-chaos behavior of certain cellular automata. (See, for ex-ample, [Langton].) Edge-of-chaos auto-mata are those whose runs become nei-ther trivially predictable nor apparently chaotic. We tend to find this sort of (emergent) behavior interesting but dif-ficult to formalize. We are not aware of any formal characterization of properties that distinguish edge-of-chaos phenom-ena from others.

To be clear, our definition of emergence allows us to label as emergent anything that is computable from a model state or sequence of states. Thus any conceptual construct that one can impose on a model counts as an emergent phenome-non.

This may seem overly simple or overly broad, and it certainly seems anti-climactic. But this definition does seem to include everything we think of as emergent. Whether it also includes phe-nomena that we would not want to con-sider emergent is open to debate.[22]

---

[22] Let's contrast our definition of emergence with Bedau's. For Bedau, a property or phenomenon is

More to the point, though, we are proposing that the essence of emergence is the independent conceptualization of the emergent phenomenon.

- Any phenomenon that may be understood in its own terms and whose understanding does not depend on knowing how it is implemented is emergent under our definition.

- No phenomenon that cannot be so understood will be emergent for us.

Emergent phenomena are often associated with a sense of surprise: where did that come from? how did that happen? Clearly the surprise of an observer has little to do with how one should understand nature. Surprise is not uncommon, however, because phenomena are typically thought of as emergent when we perceive them before we understand how they are brought about. Sometimes it may appear that there is no way for a surprising result to be brought about. Of course there always is, but this reaction is reminiscent of the remark by Arthur C. Clarke that

> any sufficiently advanced technology is indistinguishable from magic.

# 4  Implications of emergence

The technique of implementing an abstract design by using lower level constructs is not new; it is the bread and butter of Computer Science. In a recent review [CFCS] of the status of Computer Science, the Committee on the Fundamentals of Computer Science: Challenges and Opportunities, National Research Council wrote (p. 65) the following.

> [A]bstraction is a quintessential activity of computer science ... . Computer scientists create and discard abstractions as freely as engineers and architects create and discard design sketches. ...

> [S]oftware-design techniques [which allow one to retain the independence of an abstraction by maintaining the distinction between the abstraction and its implementation][23] provide ways to organize the abstract definitions and the information they control. ... [P]rogramming languages provide a notation to encode abstractions so as to allow their direct execution by computer.

This section explores the implications of emergence that our Turing Machine example illustrates.

---

emergent [Bedau] if it is applicable at some macro level but not at any micro sublevel. If one identifies *macro* with *epiphenomenal* and *micro* with *underlying*, our definition and Bedau's are quite similar except that Bedau explicitly excludes what he calls trivially emergent properties. A property is trivially emergent for Bedau if (to use our terms) it is essentially the same property at the epiphenomenal and underlying levels. As an example, Bedau's definition excludes the mass of an aggregate from being an emergent property of the aggregate because mass is also a property of the components. Our definition does not make that exclusion. For us the mass and (perhaps more interestingly) the center of gravity of an aggregate both qualify as emergent.

Our definition is similar to but less restrictive than Shalizi's [Shalizi 2001] and [Shalizi 2005], which requires that an emergent phenomenon be computationally simpler than the phenomena from which it emerges. If an epiphenomenon has its own formalizable abstraction, it will often (but even then not always) be the case that computations performed in terms of that abstraction will be simpler than the equivalent computations performed in terms of the underlying model. But we don't require this. In addition, many epiphenomena do not have easily formalizable abstractions.

---

[23]  Although the bracketed texts says a bit more than was literally in the original, we believe it expresses the authors' intentions.

## 4.1   Non-reductive regularities

Recall Weinberg's statement.

> [T]here are no autonomous laws of weather that are logically independent of the principles of physics.

Clearly there are lots of autonomous "laws" of Turing Machines (namely computability theory), and they are all logically independent of the rules of the Game of Life.

The fact that one can emulate a Turing Machine on a Game of Life platform tells us nothing about Turing Machines—other than that they can be emulated by using the Game of Life.

An emulation of a Turing Machine on the Game of Life is an example of what might be called a non-reductive regularity. The Turing Machine and its emulation is certainly a kind of regularity, but the regularity that it embodies (i.e., that it is a model for computability) is not a logical consequence of (i.e., is not reducible to and cannot be deduced from) the Game of Life rules.

Facts about Turing Machines, i.e., the theorems of computability theory, are derived *de novo*. They are made up out of whole cloth; they are not based on the Game of Life rules. The fact that such abstract designs can be realized using Game of Life rules as an implementation platform tells us nothing about computability theory that we don't already know.

## 4.2   Downward entailment

On the other hand, the fact that a Turing Machine can be implemented using the Game of Life rules as primitives does tell us something about the Game of Life—namely that the results of computability theory can be applied to the Game of Life. The property of being Turing complete applies to the Game of Life precisely because a Turing Machine can be shown to be one of its possible epiphenomena.

In other words, epiphenomena are downward entailing. Properties of epiphenomena are also properties of the phenomena from which they spring. This is not quite as striking as downward causation[24] would be, but it is a powerful intellectual tool.

Earlier, we dismissed the notion that a glider may be said to "go to a cell and turns it on." The only things that turn on Game of Life grid cells are the Game of Life rules. But because of downward entailment, there is hope for talk of this sort.

To prove that a Turing Machine emulation on a Game of Life platform does what we claim it does we disengaged the design of the emulation from its implementation, and we reasoned about the emulation as an abstract design. We can do the same thing with gliders. We can establish a domain of discourse about gliders as abstract entities. Within that domain of discourse we can reason about gliders, and in particular we can reason about which cells a glider will turn on and when it will turn them on. Our theory will tell us how fast an in which direction a glider moves

Having developed facts and rules about gliders as independent abstract elements, we can then use the fact that gliders are epiphenomena of the Game of Life and by appeal to downward entailment apply those facts and rules to the Game of Life cells that implement gliders.

---

[24]   See, for example [Emmeche] for a number of sophisticated discussions of downward causation.

This sounds more complex than it really is. What it really amounts to is that downward entailment justifies what we as human beings tend to do anyway: notice regularities in the world and then engage the world in terms of those regularities.

### 4.3  Reduction proofs

Consider how the unsolvability of the halting problem applies to the Game of Life. The fact that a Turing Machine can be implemented on a Game of Life platform means, among other things that the halting problem for the Game of Life—which we can define as determining whether a game of Life run ever reaches a stable (unchanging or repeating) configuration—is unsolvable.

When reasoning about insolvability one often talks of reducing one problem to another. In this case, because we can implement Turing Machines using the Game of Life, we know that we can reduce the halting problem for Turing Machines to the halting problem for the Game of Life: if we could solve the Game of Life halting problem, we could solve the Turing Machine halting problem. But we know that the Turing Machine halting problem is unsolvable. Therefore the Game of Life halting problem is also unsolvable. This sort of downward entailment reduction gives us a lot of intellectual leverage since it's not at all clear how difficult it would be to prove "directly" that the halting problem for the Game of Life is unsolvable.

Thus another consequence of downward entailment is that reducibility cuts both ways. One can conclude that if something is impossible at a higher level it must be impossible at the lower (implementation) level as well. But the only way to reach that conclusion is to reason about the higher level as an independent abstraction and then to reconnect that abstraction to the lower level. Logically independent higher level abstractions, i.e., functionality and design, matter on their own.

### 4.4  Downward entailment as science

A striking example of downward entailment is the kind of computation we do when computing the effect of one billiard ball on another in a Newtonian universe. It's a simple calculation involving vectors and the transfer of kinetic energy.

In truth there is no fundamental force of physics corresponding to kinetic energy. If one had to compute the consequences of a billiard ball collision in terms of quantum states and the electromagnetic force, which is the one that applies, the task would be impossibly complex. But the computation is easy to do at the epiphenomenal level of billiard balls. We know that the computation we do at the billiard ball level applies to the real world because of downward entailment: billiard balls are epiphenomena of the underlying reality. So even though it may be literally incorrect to say that a glider "turns on a Game of Life cell" or that "one billiard ball pushes another one in a particular direction," because of downward entailment this sort of conceptual shorthand is not only reasonable but essential for how we think about the world. But see the discussion of Newtonian mechanics in the next section for a somewhat stronger version of this perspective.

Downward entailment in the form of the process just sketched is, in fact, a reasonable description of how we do science: we build models, which we then apply to the world around us.

We are not saying that there are *forces* in the world that operate according to billiard ball rules or that there are *forces* in the Game of Life that operate according to glider rules. That would be downward causation, a form of strong emergence, which we have already ruled out. What we are saying is that billiard balls, gliders, Turing Machines, and their interactions can be defined in the abstract. We can reason about them as abstractions, and then through downward entailment we can apply the results of that reasoning to any implementation of those abstractions whenever the implementation preserves the assumptions required by the abstraction.

### 4.5 The reality of higher level abstractions

In "Real Patterns" [Dennett '91], Dennett argues that when compared with the work required to compute the equivalent results in terms of primitive forces, one gets a "stupendous" "scale of compression" when one adopts his notion of an intentional stance [Dennett '87]. Although "Real Patterns" doesn't spell out the link explicitly, Dennett's position appears to be that because of that intellectual advantage, one should treat the ontologies offered by the intentional stance as what he calls "mildly real"—although he doesn't spell out in any detail what regarding something as "mildly real" involves.

Our position contrasts with Dennett's in that we claim that nature is often best understood—the theories of science are best expressed—in terms of two-level (or perhaps multi-level) theories. One level is an abstract design; the other level is the implementation of that design. Whether or not one wants to say that the abstract design is "mildly real"

(or real with some other adjective applied) is not our focus.[25]

As we shall see below, our claim will also be that the entities (such as billiard balls) about which higher level abstractions are formulated are real in an objective sense (they have reduced entropy) but that interactions among those entities are epiphenomenal—since the only forces in nature are the fundamental forces.

In a recent book [Laughlin], Laughlin argues for what he calls *collective principles of organization*, which he finds to be at least as important as reductionist principles. For example in discussing Newton's laws he concludes from the fact that (p. 31)

> these [otherwise] overwhelmingly successful laws … make profoundly wrong predictions at [the quantum] scale

that

> Newton's legendary laws have turned out to be emergent. They are not fundamental at all but a consequence of the aggregation of quantum matter into macroscopic fluids and solids—a collective organizational phenomenon. … [Newton's laws] are as exact and true as anything we know in physics—yet they vanish into nothingness when examined too closely. … [M]any physicists remain in denial. To this day they organize conferences on the subject and routinely speak about Newton's laws being an "approximation" for quantum mechanics, valid when the system is large—even though no legitimate approximation scheme has ever been found.

---

[25] Furthermore, our position is that the question of whether or not anyone would find it advantageous to have beliefs about epiphenomena (which is Dennett's primary concern) has nothing to do with the issues we are considering.

A second example to which Laughlin frequently returns is the solid state of matter, which, as he points out, exhibits properties of rigidity and elasticity. The solid state of matter may be characterized as material that may be understood as a three dimensional lattice of components held together by forces acting among those components.

Once one has defined an abstract structure of this sort, one can derive properties of matter having this structure. One can do so without knowing anything more about either (a) the particular elements at the lattice nodes or (b) how the binding forces are implemented. All one needs to know are the strengths of the forces and the shape of the lattice.

From our perspective, both Newton's laws and the solid state of matter are abstract organizational designs, i.e., epiphenomena. They are abstractions that apply to nature in much the same way as a Turing Machine as an abstraction applies to certain cell configurations in the Game of Life. Laughlin calls the implementation of such an abstraction a protectorate.

Laughlin points out that protectorates tend to have feasibility ranges, which are often characterized by size, speed, and temperature. A few molecules of $H_2O$ won't have the usual properties of ice. And ice, like most solids, melts when heated to a point at which the attractive forces are no longer able to preserve the lattice configuration of the elements. Similarly Newton's laws fail at the quantum level.

The existence of such feasibility ranges does not reduce the importance of either the solid matter abstraction or the Newtonian physics abstraction. They just limit the conditions under which those abstractions apply, i.e., under which nature is able to implement them.

The more general point is that nature implements a great many such abstract designs. As is the case with computability theory, which includes many sophisticated results about the Turing machine abstraction, there are often sophisticated theories that characterize the properties of such naturally occurring abstractions. These theories may have nothing to do with how the abstract designs are implemented. They are functional theories that apply to the abstract designs themselves. To apply such theories to a real physical example (through downward entailment), all one needs is for the physical example to implement the abstract designs.

Furthermore and perhaps more importantly, these abstract designs are neither derivable from nor logical consequences of their implementations. Abstract designs and the theories built on them are new and creative constructs and are not consequences of the platform on which they are implemented. The Game of Life doesn't include the concept of a Turing machine, and quantum physics doesn't include the concept of a solid.

The point of all this is to support Laughlin position: when nature implements an abstraction, the epiphenomena described by that abstraction become just as real any other phenomena, and the abstraction that describes them is just as valid a description of that aspect of nature as any other description of any other aspect of nature.

Our notion that much of nature is best understood in terms of implementations of abstractions suggests that many scientific theories are best expressed at two levels: (1) the level of an abstraction itself, i.e., how it is specified, how it works on the abstract level, and what its implications are, and (2) the level that explains (a) under what circumstances (when and where) that abstraction may be found implemented and (b) how that implementation works.

The use of agent-based modeling in the social sciences illustrates this methodology. Agent-based models are typically used to show both (a) how a higher level abstraction functions in its context, i.e., what its consequences as a theory are, and (b) how that abstraction may come to be realized in terms of some particular set of lower level interactions.

## 4.6 Abstractions and abstract designs

We have intentionally used the terms *abstraction* and *abstract design* somewhat interchangeably. *Abstraction* is, of course, a much more general term. Yet from the perspective of scientific explanation, what one is looking for is, as Weinberg says, a description of why nature is the way it is.

To some extent, this includes the question of why nature is made up of whatever makes it up. But for the most part, this question is generally taken to be asking why nature works the way it does.

As summarized by Woodward [Woodward], scientific explanations are intended to explain

> *why* things happen, where the "things" in question can be either particular events or something more general—e.g., regularities or repeatable patterns in nature.

Typically we are looking for an operational explanation: how does nature work?[26] Thus when we speak of an abstract design, we are referring to the design of an operational mechanism that can be understood as bringing about some consequence. From here on, when we talk about *abstractions* or *abstract designs*, this is what we have in mind: the abstract design of a mechanism that produces certain kinds of results.

Note that it is not just the functional results that matter; the abstract operational design matters also. A Turing Machine is not just a device whose input-output functionality satisfies the theorems of computability theory; it is an abstract device that operates in a particular way. The theorems of computability theory follow from the way a Turing Machine operates, not the other way around. It is the abstract operational design that comes first. The theory follows.

This operational perspective seems to be somewhat different from that of traditional functionalism. Functionalism is concerned with the functions and regularities that characterize the special sciences. It tends not to be concerned with how those functions or regularities are realized. In fact the opposite seems to be the case. By emphasizing the possibility of multiple realizability, functionalism minimizes the importance of any particular realization and dismisses the importance of understanding operational design issues.

## 4.7 Phase transitions

Those of us in Computer Science know that implementations of abstract designs are often not perfect. Most include some compromises, and some have bugs.

---

[26] As Woodward explains, though, the question of what should be taken as a scientific explanation is the subject of continuing investigation.

What about nature's implementation of abstract designs?

Nature's implementation of abstract designs are not always perfect either—especially, as Laughlin points out, either (a) when the conditions under which an implementation is feasible are not in effect or (b) when one approaches the boundaries of such feasibility regions.

There will almost always be borderline situations in which the implementation of an abstract design is on the verge of breaking down. These borderline situations frequently manifest as what we call phase transitions—regions or points (related to a parameter such as size, speed, temperature, and pressure) where multiple distinct and incompatible abstractions may be implemented.

Newton's laws fail at both the quantum level and at relativistic speeds. If as Laughlin suggests, the Newtonian abstraction cannot be shown to be an approximation of quantum theory, phase transitions should appear as one approaches the quantum realm.

As explained by Sachdev [Sachdev], the transition from a Newtonian gas to a Boise-Einstein condensate (such as super-fluid liquid helium) illustrates such a phase transition.

> At room temperature, a gas such as helium consists of rapidly moving atoms, and can be visualized as classical billiard balls which collide with the walls of the container and occasionally with each other.

> As the temperature is lowered, the atoms slow down [and] their quantum-mechanical characteristics become important. Now we have to think of the atoms as occupying specific quantum states which extend across the entire volume of the container. … [I]f the atoms are 'bosons' (… as is helium) an arbitrary number of them can oc-cupy any single quantum state … If the temperature is low enough … *every* atom will occupy the *same* lowest energy … quantum state.

On the other hand, since Newton's laws are indeed an approximation of relativistic physics, there are no Newtonian-related phase transitions as one approaches relativistic speeds.

These considerations suggest that whenever data that suggests a phase transition appears, one should look for two or more abstractions with implementations having overlapping or adjacent feasibility regions.

Speculatively, one might even view quantum probability amplitude waves as phase transition phenomena. Quantum states are discrete. Matter may not occupy states that are intermediate between them. Consequently, matter cannot transition smoothly from one quantum state to another. As Hardy suggests [Hardy], by making such transitions probabilistically continuous,

> quantum theory offers us a way to have the advantages of discreteness and continuity at the same time.

Quantum waves exhibit interference patterns. It might be worthwhile to attempt to characterize non-quantum phase transitions as wave phenomena and to look for interference patterns there.

## 4.8 The constructionist hypothesis revisited

> Note to editor. This subsection may be presented as a sidebar.

Earlier we noted Anderson's rejection of what he called the *constructionist hypothesis*,

> "[that the] ability to reduce everything to simple fundamental laws … implies the ability to start from those laws and recon-
struct the universe"

Our considerations in this section clarify this rejection.

The constructionist agenda is simply not within the realm of science. Even were it theoretically possible, it simply is not one of the tasks that science sets for itself to reconstruct the universe from the principles of physics and some initial state of the universe.[27] Science simply does not attempt to imagine all the possible configurations of matter and energy that are consistent with the constraints imposed by the fundamental laws of physics, i.e., all the possible abstract designs that nature may possibly implement, and predict which will come to pass.

It is worth considering what the role of experiments would be were it possible to start with the fundamental laws of physics and reconstruct the universe. We can think of two.

1. To help refine our understanding of the fundamental laws of physics.

2. If the laws of physics require that one of a number of possible situations be the case, to decide which of them actually is the case.

Most experiments are done for neither of these reasons. Most experiments are done to help establish whether some higher level regularity actually holds.

Recently in the news were the results of the impact of a projectile with the comet Tempel 1. As reported by Cowan [Cowan], the experiment

---

revealed several surprises. The data … are at odds with a leading model for the structure of comets called the dirty-snowball model. …

[T]he data from the Deep Impact mission indicate that although Tempel 1 contains some ices, its primary constituent may be dust particles finer than talcum power. …

What's more, the comet isn't a mere hodgepodge of different materials and structures. "The damn thing is layered like a frozen onion," says Deep Impact scientist Joseph Veverka of Cornell University. …

Deep Impact's revelations "are going to change lot of our ideas about comets," predicts [Jay] Melosh [of the University of Arizona in Tucson].

These are not the words of scientists in the process of deriving results about comets from the fundamental laws of physics. These are scientists building higher level models.

It is just as wrong for Weinberg to denigrate the notion that there can be independent principles of chemistry or psychology as it would be for him to claim that the principles of Mathematics and Computer Science are all reducible to and derivable from physics.

Most sciences are like Mathematics and Computer Science. They really do stand on their own—which is why (as Anderson says) the constructionist hypothesis is wrong and why (as Fodor says) the special sciences are autonomous..

## 5  Entities

So far, we have discussed what one might characterize as emergence in the large. There is also emergence on a smaller and more local scale. That sort of emergence is related to what we intuitively think of as entities.  This section

---

[27] Although see our comments about theories of the origin of the universe and biological evolution in the final section.

discusses entities and how they relate to emergence.

We think in terms of entities, i.e., things or objects. It seems like the most natural thing in the world. Yet the question of how one might characterize what should and should not be considered an entity has long been a subject of philosophical study. A brief review of the recent literature (for example, [Boyd], [Laylock], [Miller], [Rosen], [Varzi Fall '04]) suggests that no consensus about how to understand the notion of "an entity" has yet been reached.

One might adopt a very general position. For example, Laylock quotes Lowe [Lowe] as follows.

> 'Thing', in its most general sense, is interchangeable with 'entity' or 'being' and is applicable to any item whose existence is acknowledged by a system of ontology, whether that item be particular, universal, abstract, or concrete. In this sense, not only material bodies but also properties, relations, events, numbers, sets, and propositions are—if they are acknowledged as existing—to be accounted 'things'.

For our purposes, this is too broad. In this paper we want to exclude properties, relations, events, numbers, sets and propositions from our notion of entity. We don't want to think of, say, the American Civil War or happiness as an entity in the same way that we think of an atom is an entity.

On the other hand, we don't want to limit ourselves to strictly material objects. We want to include countries, teams, corporations, and families, for example, as well as what may seem like quasi-physical entities such as people and hurricanes, whose physical makeup undergoes continual change.

For our purposes, entities, by fiat, will always have some material aspect. That is, an entity will at any time consist of physical elements arranged in a particular way. With this decision we are excluding from our notion of entity strictly mental constructs such as sets, numbers, concepts, propositions, relationships, designs, abstractions, etc.

If the preceding does not formally exclude instants, events, and durations, we will explicitly exclude them too. Entities for us will be required to persist in time, but they will not be aspects of time, i.e., instants or durations, or events, whatever an event is.

An entity for us will be either atomic (not in the sense of being a chemical element but in the more generic sense of having no constituents—if indeed there are atomic physical elements in nature), or, if an entity has constituents, it will be an epiphenomenon of its constituents. Thus for us non-atomic entities will represent one of the most common forms of emergence.

Our purpose in this section is not to settle the grand philosophical question of what one should mean by the terms *thing, object* or *entity* but to sketch out what it means to be an entity in our sense. Of course we hope that the framework we develop will offer a useful way of thinking about some of the uses to which we commonly put the terms *thing, object,* and *entity*.

### 5.1 Entities, entropy, designs, and functionality

The standard model of physics includes fundamental particles such as electrons, photons, quarks, etc. These are entities which have no constituents. Beyond these, one has atomic nuclei, atoms, and molecules, all of which we want to include in our notion of entity.

For functionalism, entities are everywhere: mice and cans are good examples. The higher level sciences speak of all sorts of entities, including biological entities (e.g., you and me) and social, political, and economic entities, such as families, states, and corporations.

We propose to characterize an entity as either atomic or as any demarcatable region that exhibits a persistent and self-perpetuating reduced level of entropy.

Since we are not prepared to define the term *demarcatable region*, perhaps defining *entity* in terms of a notion as loosely defined as demarcatable region doesn't get one very far.[28] But the notion of an entity always seems to imply a boundary that distinguishes the entity from its surroundings. Entities in our sense always have an "inside."

We discuss two kinds of entities: entities at an energy equilibrium and entities that are far from equilibrium.[29]

It is important to note that since non-atomic entities have a reduced level of entropy, they always have an internal structure, i.e., a design. Furthermore, the design of an entity often allows it to assume one or more states. A good example is the design of an atom: a nucleus along with associated electrons in various orbitals. Among the states of an atom are those differentiated by the differing energy levels of its electrons.

It seems pretty clear that we (and other animals) have evolved the ability to perceive entities in this sense. Our intuitive sense of entity seems to map fairly well onto the notion of a persistent demarcatable region that displays some special order that distinguishes it from its environment, i.e., an area that has an internal design.

We use the term *design* deliberately. Entities implement abstract designs in much the same way as abstract designs such as Newtonian mechanics are implemented on a larger scale. Because an entity implements a particular design, it exhibits the functionality that its design produces. One of the tasks of science, then, is to decide for any entity (or category of entities), what design it embodies and what the implications of that design are for the behavior of that entity (or those entities).

When nature implements an abstract design such as solid matter or Newtonian mechanics it is the functionalities that come along with that design—what the design implies about how matter that implements it behaves—that make us interested in it. These larger scale abstract designs are typically embodied by substances or by arbitrary collections of things. In contrast, the designs that entities implement produce a particular kind of functionality in a constrained and bounded region.

## 5.2  Entities at an energy equilibrium

The entities of physics and chemistry are at an energy equilibrium. A distinguishing feature of these entities is that the mass of any one of them is strictly smaller than the sum of the masses of its components. This may be seen most clearly in nuclear fission and fusion, in which one starts and ends with the same number of atomic components, i.e., electrons, protons, and neutrons—which raises the obvious question: which mass was converted to energy?

---

[28]  As Varzi [Varzi Spring '04] points out, the notion of a boundary is itself quite difficult to pin down. Some boundaries, Mt. Everest's, for example, are quite vague.

[29]  We first proposed this in [Abbott].

The answer has to do with the strong nuclear force, which implements what is called the "binding energy" of nucleons within a nucleus. Without going into details, the bottom line is that the mass of, say, a helium nucleus (also known as an alpha particle, two protons and two neutrons), which is one of the products of hydrogen fusion, is less than the sum of the masses of the protons and neutrons that make up an alpha particle when not bound together as an alpha particle.[30]

The same entity-mass relationship holds for all physical and chemical entities. The mass of an atom or molecule is (negligibly) less than the sum of the masses of its components taken separately. The mass of the solar system is (negligibly) less than the mass of the sun and the planets when taken separately.

This fact implies that the entropy of these entities is lower than the entropy of the components taken separately. In other words, an entity at an energy equilibrium is distinguishable by the fact that it has lower mass and lower entropy than its components taken separately.

These entities are trivially self-perpetuating in that they are in what is often called an energy well and require energy to pull their components apart. This gives us a nice metric of entityness for at-equilibrium entities: the amount of energy required to pull it apart.

---

[30] It turns out that the atomic nucleus with the least mass per nucleon is iron. Energy from fusion is possible for elements lighter than iron; energy from fission is possible for elements heavier than iron. (See [Nave] for a discussion of these matters.)

## 5.3 Entities and emergence are fundamental

The mechanisms (gravity, the strong nuclear force, and the electromagnetic force) that expel entropy from at-equilibrium entities and that hold these entities together are the fundamental forces of nature.

One can say that these mechanisms in some sense run for free. To the extent that we understand how they work at all, we attribute their operation to virtual particles that pop into and out of existence and that do the work of the force—with no extra effort expended anywhere else.

There really is a free lunch. Atomic nuclei form, atoms form, solar systems and galaxies form—all without depleting any energy reservoirs. We are so used to this fact that we hardly notice it. But if one stands back and observes that at-equilibrium entities exemplify emergence at its most basic—an atom is emergent from, it is an epiphenomenon of, and it supervenes over its components—we may conclude that spontaneous emergence is fundamental to how nature works.

Even so, one might suppose that beyond combining in these basic ways (as atomic nuclei, atoms, and astronomical aggregations held together by gravity), at-equilibrium entities are not very interesting. Standing back again makes it clear that this is not the case. Given what we have learned during the past half century (and what we still don't know)—especially about condensed matter physics and including, as we said earlier, the startling fact that the same matter is capable of implementing multiple abstractions with radically different properties—at-equilibrium entities are far from boring.

## 5.4 Dissipative structures

In [Prigogine] (and elsewhere) Prigogine discussed what he called a dissipative structure. We see dissipative structures as the essential stepping stone from at-equilibrium entities to autonomous entities.

Intuitively, a dissipative structure typically manifests when energy is pumped into a bounded region. Dissipative structures typically involve structured activities internal to the region. A standard example consists of the Bénard convection cycles that form in a liquid when one surface is heated and the opposite surface is kept cool. (See Figure 4.)

A number of interesting phenomena may be understood as dissipative structures. Consider the distribution of water over the earth. Water is transported from place to place via processes that include evaporation, atmospheric weather system movements, precipitation, groundwater flows, ocean current flows, etc. Taken as a global system, these cycles may be understood as a dissipative structure that is driven primarily by solar energy, which is pumped into the earth's atmosphere and surface structures. All of this is played out against a static framework defined and held in place by the earth's surface and its gravitational field.

We note that our definition of a dissipative structure is quite broad. It includes virtually any energy-consuming device that operates according to some design. Consider a digital watch. It converts an inflow of energy into an ongoing series of structured internal activities. Does a digital watch define a dissipative structure? One may argue that the design of a digital watch limits the ways in which it can respond to an energy inflow. Therefore the structured activity that arises as energy is pumped into it should not be characterized as a dissipative structure.

But any bounded region has only a limited number of ways in which it can respond to an inflow of energy. We suggest that it would be difficult if not impossible to formalize a principled distinction between the Bénard convection cycles that arise in a liquid when energy is pumped into it and the structured activities within a digital watch.[31] The primary difference seems to be that a digital watch has a much more constrained static structure and can respond in far fewer ways.

Recall that we previously characterized Newtonian mechanics and the solid phase of matter as abstractions that matter implements under various conditions. We can do the same thing for dissipative structures and say that a dissipative structure appears within a bounded region when the materials within that region implement an energy-driven abstract design.

An apparent difference between the abstract designs that dissipative structures implement and the abstract designs discussed earlier is that the abstract designs of dissipative structures seem to appear unbidden—we don't expect them—whereas the abstract designs discussed earlier are commonplace. The issue for the more commonplace abstract designs is how to conceptualize them, not why they appeared at all, whereas the abstract design that appear as dissipative structures seem to demand an answer to the question: why did they appear at all? In fact, both kinds of abstract design are

---

[31] One of the other common examples of a dissipative structure is the Belousov-Zhabotinsky (BZ) reaction, which in some ways is a chemical watch.

part of nature. The difference is that some are familiar; others aren't.[32]

If we understand a dissipative structure to be the implementation of an energy-driven abstract design, the question for any dissipative structure becomes: what abstract design does it implement? In other words, how does it work—which is the same question one must ask about any abstract design.

Like most abstract designs, those associated with dissipative structures generally exist only within limited energy ranges. Thus phase transitions may be expected as materials transform themselves between configurations in which they are and are not implanting the abstract design of a particular dissipative structure.

In this section we have referred, somewhat awkwardly, to bounded regions within which dissipative structures form. We have refrained from calling these bounded regions entities. This may be pickiness on our part, but our notion is that an entity perpetuates itself. As defined, bounded regions of materials that are capable of implementing dissipative structure abstract designs need not have the capacity to perpetuate themselves. Their boundaries may be imposed artificially. We shall have more to say about this in the section on natural vs. artificial autonomous entities.

---

[32] Reliance on a surprise factor as characteristic of emergence is, in our opinion, a common error. One often hears that emergent phenomena appear unbidden, that one's surprise is fundamental to whether something should be considered emergent. On the contrary, we claim that an observer's surprise of lack of surprise should have nothing to do with how we should understand a phenomenon of nature.

## 5.5 Integrating dissipative structures and at-equilibrium entities.

A dissipative structure is a physical manifestation of a region of energy stability in an environment in which energy is flowing at a relatively constant rate. An at-equilibrium entity is similarly a physical manifestation of a region of stability—but in an environment in which there is no energy flow. We would welcome a formal integration of the two in which at-equilibrium entities are understood as dissipative structures in an environment in which the rate of energy flow is zero. Perhaps another way of putting this would be to characterize the energy wells that exist in environments that include energy flows.

## 5.6 Autonomous entities

The notion of an autonomous entity seems central to how we look at the world.

- For millennia we have found it convenient to partition the world into two realms: the animate and the inanimate. The inanimate world is ruled by external forces; the animate world is capable of autonomous action. Recall that this is why Brownian motion posed such a problem: how can inanimate particles look so much like they are moving autonomously?

- For the past half-millennium western civilization (and more recently civilization world-wide) has pursued, with significant success, the dream of creating autonomous sources of action. We have built machines about which it can be said that in varying degrees they act on their own. We do not yet confuse our machines with biological life, and we have not yet managed to construct biological life "from scratch." But the differences between human artifacts and natural biological life are becoming more and more subtle—and they are likely to disappear within the lifetimes of many of us.

• Most people will acknowledge that the kinds of entities that the biological and social sciences deal with seem somehow different from those of physics and chemistry. A major part of that difference is the apparent ability of the entities in those sciences to act on their own, i.e., their autonomy.

So, what do we mean by autonomy? Certainly, we no longer believe in anything like vitalism, i.e., that there is such a thing as a "life force" the possession of which differentiates the animate from the inanimate. But when we speak of autonomous entities, have we done much more than substitute the word *autonomous* for other words? Do we have a serviceable definition of what it means to be autonomous?

In non-political contexts, the term *autonomous* is generally taken to mean something like *self-directed* or *not controlled by outside forces*.[33] But definitions of this sort don't help much. Perhaps *self-directed* is what we mean by *autonomous*. But what do we mean by *self-directed*?

Furthermore any entity (in our sense of an entity as having some material aspect) is subject to outside, i.e., physical, forces. Nothing is free from the laws of physics. So it may not make any sense to demand that to be autonomous an entity must not be controlled by outside forces.

The intuition behind self-directed and the connection to outside forces may give us a clue, however. Perhaps one can require that an autonomous entity control—at least to some extent and in what may be considered a self-directed way, although without implying willfulness—*how* it is affected by outside forces.

Putting these ideas together, we suggest that a useful way to think about autonomy may be that an entity is autonomous to the extent that it shapes the way it is affected by outside forces.

But this is pretty much how we have defined a dissipative structure. A dissipative structure results from the operation of an energy-driven abstract design. In other words, a dissipative structure results when an energy-driven abstract design shapes the way outside forces operate within a bounded region.

Because this seems to be such a nice fit with our intuition of what it means for an entity to be autonomous, we will define an autonomous entity as an entity that is implementing the abstract design of a dissipative structure.[34]

In other words, we define an autonomous entity as a self-perpetuating region of reduced entropy that is implementing a dissipative structure's abstract design. By definition, autonomous entities consume energy and are far from equilibrium. We suggest that most if not all of the entities of the higher level sciences satisfy our definition of an autonomous entity.

Note that most biological, social, and economic autonomous entities are even more autonomous than our definition suggests. Most of these entities acquire energy in some "frozen" form such as food or money[35] and convert it to energy according to their internal designs. Thus they do more than simply shape how "raw" energy that they encounter affects them. They are often able to save energy and to chose in some sense when to use it.

---

[33] See, for example, the American Heritage® Dictionary definition. URL as of 9/15/2005: http://www.bartleby.com/61/86/A0538600.html.

[34] This intuitive fit may be one reason that the notion of a dissipative structure generated as much enthusiasm as it has.

### 5.7 A naturally occurring autonomous entity that is neither biological nor social

We suggest that a hurricane qualifies as an autonomous entity. (See Figure 3.) In simple terms (paraphrased from [NASA]), the internal design of a hurricane involves a greater than normal pressure differential between the ocean surface and the upper atmosphere. That pressure differential causes moist surface air to rise. When the moisture-laden air reaches the upper atmosphere, which is cooler, it condenses, releasing heat. The heat warms the air and reduces the pressure, thereby maintaining the pressure differential—a marvelous design for a self-perpetuating process.

In effect, a hurricane is a heat engine in which condensation, which replaces combustion as the source of heat, occurs in the upper atmosphere.[36] Thus, although physically very large, a hurricane has a relatively simple design, which causes it to consume energy and which allows it to perpetuate itself as an area of reduced entropy.

### 5.8 Natural and artificial autonomous entities

Most of our energy consuming machines also qualify as autonomous entities. The primary difference between human pro-duced autonomous entities and naturally occurring ones is that the naturally occurring autonomous entities use at least some of the energy they consume to perpetuate themselves as entities. In contrast, human-produced autonomous entities are almost always at-equilibrium entities through which energy flows. In other words, the nature of human-produced autonomous entities is that their persistence as entities tends to be independent of their use of the energy that flows through them. This tends not to be the case with naturally occurring autonomous entities.

One of the senses of the word *natural* is to have properties characteristic of elements found in nature. We suggest that the distinction between entities that rely on an at-equilibrium frame and those that more actively construct their framework is one of the central intuitive differences between what we call artificial and what we call natural. A hurricane would thus be considered a naturally occurring autonomous entity which is neither biological nor social.

As an example of a naturally occurring at-equilibrium entity that becomes autonomous, consider an atom that is being excited by a photon stream. Because of its design it captures the energy of the photons, which it releases at some later time in what may be a slightly different form. This is the basis of the laser.

### 5.9 Autonomous entities and phase transitions

Many autonomous entities exhibit the equivalent of phases—and phase transitions. Such phases differ from phases in at-equilibrium entities in that they reflect different ways in which the autonomous entity makes use of the energy that is flowing through it. Examples include gaits (walking, running, etc.), heart beats (regular and fibrillation), and possibly psychological conditions such as mania, depression and psychosis.

---

[35]  The maxim *follow the money* is really advising to follow the energy.

[36]  A characterization of hurricanes as "vertical heat engines" may be found in Wikipedia. URL as of 9/1/2005: http://en.wikipedia.org/wiki/Hurricane

The primary concern about global warming is not that the temperature will rise by a degree or two—although the melting of the ice caps resulting from that is potentially destructive—but the possibility that if the temperature warms sufficiently, a phase transition will occur, and the global climate structure, including atmospheric and oceanic currents, will change abruptly—and possibly disastrously.

As we suggest later, the fact that parallels exist between autonomous and at-equilibrium entities leads to the suggestion that one might be able to integrate the two and see at-equilibrium entities as one end of a continuum that includes both at-equilibrium and autonomous entities.

### 5.10 Autonomous entities and energy flows

Autonomous entities require energy flows for survival. But the kinds of energy flows available are limited. The most familiar (at least here on earth) is the flow of energy from the sun. Plants exploit it. We are also familiar with artificial energy flows, as in the flow of electricity to a device when the switch is turned on. Other than these, what other flows of energy support autonomous entities?

Thermal vents in the ocean are one possibility. Yet the primary food producers in thermal vents are bacteria that convert chemicals from the vents to more useable forms of energy.[37] It is not clear what role, if any, is played by the flow of thermal energy itself. It would be significant if a life-form were found that used thermal energy directly to power an internal process in a way that paralleled the way plants use energy from the sun. It may be that some of the chemical reactions that occur in inhabitants of vent ecologies depend on a high ambient temperature. But that seems to be a different sort of dependency than using a direct energy flow.

---

[37] See, for example, Comm Tech Lab and University of Delaware.

Most biological autonomous entities acquire their energy in a packaged form, e.g., as "food" of some sort rather than as a direct energy flow. Once the energy resource has been ingested, energy is extracted from it. This is even the case with our hurricane example. The energy of condensation is produced within the hurricane after warm moist air is "ingested."

This seems to be another distinction between naturally occurring and artificial autonomous entities. No artificial entities procure their own energy resources. Other than plants, all naturally occurring autonomous entities do.

### 5.11 Theseus's ship

The distinction between natural and artificial entities sheds some light on the paradox of Theseus's ship, a ship that was maintained (repaired, repainted, etc.) in a harbor for so long that all of its original material had been replaced. Does one say that it is "the same ship" from year to year?

We would like to distinguish between two ways of looking at Theseus's ship. One way is to consider the material ship as it exists at any one moment. By our definition, this is an entity—although it is not an autonomous entity—since it is at an energy equilibrium. It is held together by a large number of relatively shallow energy wells. Entities of this sort are particularly vulnerable to every-day weathering and wear and tear. It doesn't take much to push some of the energy wells beyond their limits.

A second way to look at Theseus's ship is to include the maintenance process as part of a larger autonomous ship entity. The ship along with its maintenance process is an entity because it is a self-perpetuating region of reduced entropy. It is a relatively simple example of a social autonomous entity. Both materials and people cycle through it, but the process perpetuates itself by using energy from the society in which it is embedded.

So our answer to the question of whether "the same ship" is in the harbor from year to year is "No" if we are thinking about the material ship and "Yes" if we are thinking about the larger ship-plus-maintenance entity.

By our definition, the larger ship-plus-maintenance entity would be considered natural rather than artificial because it as a social process and is not at-equilibrium; it uses some of the energy it consumes to perpetuate itself. We would consider most social entities to be natural in this sense even though they are constructed and maintained by people.

### 5.12 Autonomous entities may act in the world

As we know, hurricanes can cause significant damage. So far we haven't talked about how that might happen.

Since energy flows through autonomous entities, part of that flowing through involves flowing out. In other words, autonomous entities may include as part of their designs means for projecting force into the world by directing outward flows of energy.[38]

Furthermore, the internal design of most autonomous entities enable them (a) to store energy, (b) to move it about internally, and (c) to tap it as needed.

### 5.13 Autonomous entities tend not to supervene over their static components

As we said earlier, an at-equilibrium entity consists of a fixed collection of component elements over which it supervenes. In contrast, autonomous entities for the most part tend not to consist of a fixed collection of matter. Our hurricane is a good example. A hurricane may be relatively stable as a reduced entropy region—even though its boundaries may be somewhat vague. But however its boundaries are defined, the material within its boundaries tends to vary from moment to moment as the hurricane's winds move air and water about.

Similarly, most biological entities recycle their physical components, and most social entities (e.g., families) and economic entities (e.g., corporations) remain intact as the people who fill various roles cycle through them. Theseus's ship—when understood as including its maintenance process as discussed above—is another example of an autonomous entity that recycles its physical components.

Because of this recycling property, most autonomous entities don't supervene

---

[38] This solves a problem that concerned Leibniz with respect to monads: how do they interact. Leibniz's answer was that they don't. Our autonomous entities interact with each other and with the rest of the world though energy flows over which they have the ability to exert some control. Of course our autonomous entities can exert that control because they have internal designs; Leibniz's monads didn't.

over any collection of matter that gives us any intellectual leverage.

It is easiest to see this when we consider gliders in the Game of Life, about which this is true as well. In the Appendix we show how to formalize the notion of a Game of Life pattern. In simplest terms we define what we call a *live cell group* to be a connected group of live (i.e., "on") cells. We define a pattern as a connected sequence of live cell groups. In general, such sequences may branch or terminate, but the glider pattern is a linear sequence of live cell groups. (See the Appendix for the details, which pretty much match one's intuition.) A glider is such a pattern.

One may define the state of a glider pattern to be the particular configuration it is in (See Figure 2 earlier for the four possible configurations.) Alternatively, one may also define the state of a glider pattern in either of two ways: the configuration (of the four) in which the pattern exists or the configuration along with the pattern's location on the grid.

To satisfy supervenience, for a glider pattern to supervene over a set of Game of Life cells requires that if the glider is in different states then the grid cells must also be in a different state.

Given either of our two definitions of state, gliders (if undisturbed) do not supervene over any finite set of grid cells. Given any such finite set of cells, a glider may assume multiple states when beyond that set, thereby violating supervenience.

The only sets of cells over which a glider supervenes is a superset of (an infinite subset of cells within) what one might call the glider's "glide path," the strip of cells that a glider will traverse if undisturbed. The parenthetical qualification allows for the possibility that one can differentiate states without looking at the entire glider pattern.

In other words, any set of cells over which a glider supervenes must include a potentially infinite subset of the cells with which the glider comes in contact over its lifetime. This may be supervenience, but it is supervenience in a not very useful way.

To connect this to autonomous entities, imagine a glider pattern as fixed with the grid moving underneath it, i.e., as if the glider cycles grid cells through itself. This is quite similar to how most autonomous entities operate. These entities typically cycle matter through themselves. The same reasoning shows that such autonomous entities don't supervene over any useful subset of matter other than the collection of all mater with which they may come in contact during their lifetimes.

It appears that the concept of supervenience may not be as useful as one might have hoped for thinking about epiphenomena and emergence—at least in the case of autonomous entities.

### 5.14 Entities, objects, and agents

Computer Science has also developed a distinction between entities that do and do not act autonomously. Recall that our definition of entity depended on distinguishing an entity from its environment, i.e., it was a region of reduced entropy. We may therefore refer to the "inside" of an entity and to whatever internal structure and state it may have. This also allows us also to speak of the interface (boundary) between an entity and its environment.

If an entity has an internal state, what, if anything, may cause that state to change? Are there outside influences that may cause an entity to change state? If so, what mechanism enables those influences to act on the entity? Alternatively, may an entity change state as a result of purely internal activity?

In Computer Science two concepts have emerged as fundamental to these issues: objects and agents. There is a reasonable consensus in Computer Science about what we mean by an object, namely an encapsulation of a mechanisms for assuming and changing states along with means for acting on that encapsulated mechanism.

There is far less agreement about the notion of an agent. For our purposes, we will construe an agent as simply as possible. An agent for us will be an object (as defined above) that may act on its own. In software terms, this means that an agent is a software object that has an internal thread.[39]

Given these definitions of *object* and *agent*, we suggest that to a first very rough approximation[40] objects are the software equivalent of at-equilibrium entities and agents are the software equivalent of autonomous entities.

### 5.15 Thermodynamic computing: nihil ex nihilo

Note that when discussing software objects and agents, there is no concern with entropy: the software system maintains the integrity (and internal structure) of objects and agents. Similarly, we did not claim that gliders or Turing Machines were entities in the Game of Life.

---

[39]    In adopting this definition, we are deliberately bypassing issues of goals, beliefs, plans, etc., which appear in some formulations of agent-based modeling frameworks.

[40]    See the next section for a discussion of why this approximation is indeed very rough.

The problem has to do with the way we do Computer Science. In Computer Science we assume that one can specify a Turing Machine, a Finite State Automaton, a Cellular Automaton, or a piece of software, and it will do its thing—for free. Software runs for free. Turing machines run for free. Cellular Automata run for free. Gliders run for free. Agents in agent-based models run for free. Although that may be a useful abstraction, we should recognize that we are leaving out something important. In the real world one needs energy to drive processes. To run real software in the real world requires a real computer, which uses real energy. We suggest that a theory of thermodynamic computation is needed to integrate the notions of energy, entities, and computing.

How do we capture the notion of the "energy" that enables software to do its "symbolic work?" Is computational complexity an equivalent concept? Computational complexity is concerned primarily with finding measures for how intrinsically difficult particular kinds of computations are. The focus seems different.

Performance analysis is somewhat closer to what we are attempting to get at. But performance analysis is typically satisfied with relatively gross results, not with the fine details of how a computational energy budget is spent.

The problem seems to be that the computational energy that software uses is not visible to the software itself. Software does not have to pay its energy bill; the rest of nature does.

However this issue is resolved, for now a thread seems to be a useful software analog for the energy flow that powers a dissipative structure. It also seems rea-sonable to use the term *agent* as synonymous with *autonomous entity*.

With this in mind, though, we should point out that the parallel between objects and agents on the one hand and at-equilibrium and autonomous entities on the other isn't perfect. An object in software is not completely controlled by external forces. An object's methods do shape how energy (in the form of threads that execute them) affects the object.

Objects differ from agents in that they don't have what might be considered an internal source of energy. Agents do. But our analogy breaks down entirely if an object is allowed to create a thread when one of its methods is executed. (Most multi-threaded programming languages allow the arbitrary creation of threads.) For an object to create a thread would be equivalent to an entity in nature creating an unlimited internal source of energy for itself once it came in contact with any external energy at all.

As we said, a real theory of thermodynamic computing is needed.

### 5.16 Minimal autonomous entities
In [Kauffman] Kauffman asks what the basic characteristics are of what he (also) calls autonomous agents. He suggests that the ability to perform a thermodynamic (Carnot engine) work cycle is fundamental.

In what may turn out to be the same answer we suggest looking for the minimal biological organism that perpetuates itself by consuming energy. Bacteria seem to be too complex. Viruses[41] and prions don't consume energy.[42] Is there anything in between? We suggest that such a minimal autonomous entity may help us understand the yet-to-be-discovered transition from the inanimate to the animate.

Since self-perpetuation does not imply reproduction (as hurricanes illustrate), simple self-perpetuating organisms may not be able to reproduce. That means that if they are to exist, it must be relatively easy for them to come into being directly from inorganic materials. Similarly, simple self-perpetuating organisms may not include any stable internal record—like DNA—of their design (as hurricanes again illustrate). One wouldn't expect to see evolution among such organisms—at least not evolution that depends on modifications of such design descriptions..

## 6  The evolution of complexity

### 6.1  Stigmergy
Once one has autonomous entities (or agents) that persist in their environment, the ways in which complexity can develop grows explosively.   Prior to agents, to get something new, one had to build it as a layer on top of some existing substrate. As we have seen, nature has found a number of amazing abstractions along with some often surprising ways to implement them. Nonetheless, this construction mechanism is relatively ponderous. Layered hierarchies of abstractions are powerful, but they are not what one might characterize as lightweight or responsive to change. Agents change all that.

---

[41] Viruses are an interesting contrast to our lactose example, however. In both cases, an at-equilibrium element in the environment triggers a process in an autonomous entity. In the case of lactose, the process is advantageous to the entity; in the case of viruses, it is not advantageous to the entity.

---

[42] Hurricanes aren't biological.

Half a century ago, Pierre-Paul Grasse invented [Grasse] the term *stigmergy* to help describe how social insect societies function. The basic insight is that when the behavior of an entity depends to at least some extent on the state of its environment, it is possible to modify that entity's behavior by changing the state of the environment. Grasse used the term "stigmergy" for this sort of indirect communication and control. This sort of interplay between agents and their environment often produces epiphenomenal effects that are useful to the agents. Often those effects may be understood in terms of formal abstractions. Sometimes it is easier to understand them less formally.

Two of the most widely cited examples of stigmergic interaction are ant foraging and bird flocking. In ant foraging, ants that have found a food source leave pheromone markers that other ants use to make their way to that food source. In bird flocking, each bird determines how it will move at least in part by noting the positions and velocities of its neighboring birds.

The resulting epiphenomena are that food is gathered and flocks form. Presumably these epiphenomena could be formalized in terms of abstract effects that obeyed a formal set of rules—in the same way that the rules for gliders and Turing Machines can abstracted away from their implementation by Game of Life rules. But often the effort required to generate such abstract theories doesn't seem worth the effort—as long as the results are what one wants.

Here are some additional examples of stigmergy.

- When buyers and sellers interact in a market, one gets market epiphenomena. Economics attempts to for-

malize how those interactions may be abstracted into theories.

- We often find that laws, rules, and regulations have both intended and unintended consequences. In this case the laws, rules, and regulations serve as the environment within which agents act. As the environment changes, so does the behavior of the agents.

- Both sides of the evo-devo (evolution-development) synthesis [Carroll] exhibit stigmergic emergence. On the "evo" side, species create environmental effects for each other as do sexes within species.

- The "devo" side is even more stigmergic. Genes, the switches that control gene expression, and the proteins that genes produce when expressed all have environmental effects on each other.

- Interestingly enough, the existence of gene switches was discovered in the investigation of another stigmergic phenomenon. Certain bacteria generate an enzyme to digest lactose, but they do it only when lactose is present. How do the bacteria "know" when to generate the enzyme?

It turns out to be simple. The gene for the enzyme exists in the bacteria, but its expression is normally blocked by a protein that is attached to the DNA sequence just before the enzyme gene. This is called a gene expression switch.

When lactose is in the environment, it infuses into the body of the bacteria and binds to the protein that blocks the expression of the gene. This causes the protein to detach from the DNA thereby "turning on" the gene and allowing it to be expressed.

The lactose enzyme switch is a lovely illustration of stigmergic design. As we described the mechanism above, it seems that lactose itself turns on the switch that causes the lactose-digesting enzyme to be produced. If one were thinking about the design of such a system, one might imagine that the lactose had been designed so that it would bind to that switch. But of course, lactose wasn't "designed" to do that. It existed prior to the switch. The bacteria evolved a switch that lactose would bind to. So the lactose must be understood as being part of the environment to which the bacteria adapted by evolving a switch to which lactose would bind. How clever; how simple; how stigmergic!

- Cellular automata operate stigmergically. Each cell serves as an environment for its neighbors. As we have seen, epiphenomena may include gliders and Turing Machines.

- Even the operation of the Turing Machine as an abstraction may be understood stigmergically. The head of a Turing Machine (the equivalent of an autonomous agent) consults the tape, which serves as its environment, to determine how to act. By writing on the tape, it leaves markers in its environment to which it may return—not unlike the way foraging ants leave pheromone markers in their environment. When the head returns to a marker, that marker helps the head determine how to act at that later time.

- In fact, one may understand all computations as being stigmergic with respect to a computer's instruction execution cycle. Consider the following familiar code fragment.

```
temp := x;
x    := y;
y    := temp;
```

The epiphenomenal result is that x and y are exchanged. But this result is not a consequence of any one statement. It is an epiphenomenon of the three statements being executed in sequence by a computer's instruction execution cycle.

Just as there in nothing in the rules of the Game of Life about gliders, there is nothing in a computer's instruction execution cycle about exchanging the values of x and y—or about any other algorithm that software implements. Those effects are all epiphenomenal.

- The instruction execution cycle itself is epiphenomenal over the flow of electrons through gates—which knows no more about the instruction execution cycle than the instruction execution cycle knows about algorithms.

In all of the preceding examples it is relatively easy to identify the agent(s), the environment, and the resulting epiphenomena.

## 6.2   Design and evolution

It is not surprising that designs appear in nature. It is almost tautologous to say that those things whose designs work in the environments in which they find themselves will persist in those environments. This is a simpler (and more accurate) way of saying that it is the fit—entities with designs that *fit* their environment—that survive.

## 6.3   The accretion of complexity

An entity that suits its environment persists in that environment. But anything that persists in an environment by that very fact changes that environment for everything else. This phenomenon is commonly referred to as an ever changing fitness landscape.

What has been less widely noted in the complexity literature is that when something is added to an environment it may enable something else to be added latter—something that could not have existed in that environment prior to the earlier addition.

This is an extension of notions from ecology, biology, and the social sciences. A term for this phenomenon from the ecology literature, is *succession*. (See, for example, [Trani].) Historically *succession* has been taken to refer to a fairly rigid sequence of communities of species, generally leading to what is called a climax or (less dramatically) a steady state.

Our notion is closer to that of *bricolage*, a notion that originated with the structuralism movement of the early 20th century [Wiener] and which is now used in both biology and the social sciences. *Bricolage* means the act or result of tinkering, improvising, or building something out of what is at hand.

In genetics *bricolage* refers to the evolutionary process as one that tinkers with an existing genome to produce something new. [Church].

John Seely Brown, former chief scientist for the Xerox Corporation and former director of the Xerox Palo Alto Research Center captured its sense in a recent talk.

> [W]ith *bricolage* you appropriate something. That means you bring it into your space, you tinker with it, and you repur-

> pose it and reposition it. When you repurpose something, it is yours.[43]

Ciborra [Ciborra] uses *bricolage* to characterize the way that organizations tailor their information systems to their changing needs through continual tinkering.

This notion of building one thing upon another applies to our framework in that anything that persists in an environment changes that environment for everything else. The Internet provides many interesting illustrations.

- Because the Internet exists at all, access to a very large pool of people is available. This enabled the development of websites such as eBay.

- The establishment of eBay as a persistent feature of the Internet environment enabled the development of enterprises whose only sales outlet was eBay. These are enterprises with neither brick and mortar nor web storefronts. The only place they sell is on eBay. This is a nice example of ecological succession.

- At the same time—and again because the Internet provides access to a very large number of people—other organizations were able to establish what are known as massively multi-player online games. Each of these games is a simulated world in which participants interact with the game environment and with each other. In most of these games, participants seek to acquire virtual game resources, such as magic swords. Often it takes a fair amount of

---

[43] In passing, Brown claims that this is how most new technology develops.

> [T]hat is the way we build almost all technology today, even though my lawyers don't want to hear about it. We borrow things; we tinker with them; we modify them; we join them; we build stuff.

time, effort, and skill to acquire such resources.

- The existence of all of these factors resulted, though a creative leap, in an eBay market in which players sold virtual game assets for real money. This market has become so large that there are now websites dedicated exclusively to trading in virtual game assets. [Wallace]

- BBC News reported [BBC] that there are companies that hire low-wage Mexican and Chinese teenagers to earn virtual assets, which are then sold in these markets. How long will it be before a full-fledged economy develops around these assets? There may be brokers and retailers who buy and sell these assets for their own accounts even though they do not intend to play the game. (Perhaps they already exist.) Someone may develop a service that tracks the prices of these assets. Perhaps futures and options markets will develop along with the inevitable investment advisors.

The point is that once something fits well enough into its environment to persist it adds itself to the environment for everything else. This creates additional possibilities and a world with ever increasing complexity.

In each of the examples mentioned above, one can identify what we have been calling an autonomous entity. In most cases, these entities are self-perpetuating in that the amount of money they extract from the environment (by selling either products, services, or advertising) is more than enough to pay for the resources needed to keep it in existence.

In other cases, some Internet entities run on time and effort contributed by volun-teers. But the effect is the same. As long as an entity is self-perpetuating, it becomes part of the environment and can serve as the basis for the development of additional entities.

### 6.4 Increasing complexity increasing efficiency, and historical contingency

The phenomenon whereby new entities are built on top of existing entities is now so widespread and commonplace that it may seem gratuitous even to comment on it. But it is an important phenomenon, and one that has not received the attention it deserves.

Easy though this phenomenon is to understand once one sees it, it is not trivial. After all, the second law of thermodynamics tells us that overall entropy increases and complexity diminishes. Yet we see complexity, both natural and man made, continually increasing. For the most part, this increasing complexity consists of the development of new autonomous entities, entities that implement the abstract designs of dissipative structures.

This does not contradict the Second Law. Each autonomous entity maintains its own internally reduced entropy by using energy imported from the environment to export entropy to the environment. Overall entropy increases. Such a process works only in an environment that itself receives energy from outside itself. Within such an environment, complexity increases.

Progress in science and technology and the bountifulness of the marketplace all exemplify this pattern of increasing complexity. One might refer to this kind of pattern as a meta-epiphenomenon since it is an epiphenomenon of the process that creates epiphenomena.

This creative process also tends to exhibit a second meta-epiphenomenon. Overall energy utilization becomes continually more efficient. As new autonomous entities find ways to use previously unused or under-used energy flows (or forms of energy flows that had not existed until some newly created autonomous entity generated them, perhaps as a waste product), more of the energy available to the system as a whole is put to use.

The process whereby new autonomous entities come into existence and perpetuate themselves is non-reductive. It is creative, contingent, and almost entirely a sequence of historical accidents. As they say, history is just one damn thing after another—to which we add, and nature is a bricolage. We repeat the observation Anderson made more than three decades ago.

> The ability to reduce everything to simple fundamental laws [does not imply] the ability to start from those laws and reconstruct the universe.

# 7 Entities, emergence, and science

## 7.1 Entities and the sciences

One reason that the sciences at levels higher than physics and chemistry seem somehow softer than physics and chemistry is that they work with autonomous entities, entities that for the most part do not supervene over any conveniently compact collection of matter. Entities in physics and chemistry are satisfyingly solid—or at least they seemed to be before quantum theory. In contrast, the entities of the higher level sciences are not defined in terms of material boundaries. These entities don't exist as stable clumps of matter; it's hard to hold them completely in one's hand—or in the grip of an instrument.

The entities of the special sciences are objectively real—there is some objective measure (their reduced entropy relative to their environment) by which they qualify as entities. But as we saw earlier, the processes through which these entities interact and by means of which they perpetuate themselves are epiphenomenal. Even though the activities of higher level entities may be described in terms that are independent of the forces that produce them (recall that this is our definition of epiphenomenal), the fundamental forces of physics are the only forces in nature. There is no strong emergence. All other force-like effects are epiphenomenal.

Consequently we find ourselves in the position of claiming that the higher level sciences study epiphenomenal interactions among real if often somewhat ethereal entities.

## 7.2 Science and emergence

The idea that one can use heat and the expansion of gases that it produces to implement a particular function is not a concept of fundamental physics. Of course the Carnot engine is a consequence of fundamental physics, but it is not a concept of fundamental physics. The idea of using a force to implement new functionality is simply not within the realm of fundamental physics.

Physics, like most science, does not consider new functionality. It examines existing phenomena, and it asks how they are brought about. It does not ask how knowledge gained from such an analysis can be used to implement something new.

Here is a representative definition of the term *science*.

- The observation, identification, description, experimental investigation, and theoretical explanation of phenomena. [American Heritage]

Science is thus the study of nature, how it is designed, i.e., organized, and how its designs work. Science does not have as part of its charter to take what is known about nature and to create something new.

Recall our discussion of hurricanes. Apparently they are the only kind of weather system with an internal power plant. Let's imagine that no hurricane ever existed—at least not anywhere that an earthbound scientist could observe it. Under those circumstances no scientist would hypothesize the possibility of such a weather system. Doing so just isn't part of the scientific agenda; it is not the kind of task that scientists set for themselves.

Why waste one's time thinking about something so strange—a weather system that not only contains its own built-in power plant but one in which the heat is generated by condensation rather than combustion and the "furnace" in which the heat is generated is located in the upper atmosphere. Thinking through such a possibility might make interesting science fiction.

> In a galaxy far away, on a planet of a medium size star near the edge of that galaxy, a planet that had storms with their own built-in heat engines, ....

Certainly nothing so bizarre could ever occur naturally. It would not be considered science.

Imagine also how bizarre phase transitions would seem if they weren't so common—matter sometimes obeying one set of rules and sometimes obeying another set. It wouldn't make any sense.

What would happen at the boundaries? How would transitions occur? If phase transitions didn't happen naturally, science almost certainly wouldn't invent them.

If we conceive of science as the study of existing phenomena, science *is* reductionism. To paraphrase Weinberg, the goal of science is

> to find simple universal laws that explain why nature is the way it is.

When science is understood in this way, mathematics, computer science, and engineering, all of which create and study conceptual structures that need not exist, are not science. Indeed scientists and mathematicians are often surprised when they find that a mathematical construct that had been studied simply because it seemed mathematically interesting has a scientific application.

Fortunately for us, nature is not a scientist. Like computer scientists and engineers, she too creates things that need not exist—people and hurricanes, for example.

What about this paper? We would categorize this paper as science because one of its goals is to help explain, i.e., to provide some intellectual leverage for understanding, why the nature is the way it is.

This immediately raises another question: if this is science, are we happy with it? Let's assume that the simplest and most universal way to understand nature is in terms of multilevel abstractions. Is this satisfactory? Is this approach to scientific explanation as real and as concrete as explaining nature in terms of more absolute single-level laws? Isn't there something unreal about explaining nature at least in part as implementations of abstractions?

One way to argue for the reality of these abstractions is to show that they build upon each other. When a new abstraction is implemented in terms of the functionalities embodied in existing abstractions, there seems little choice but to acknowledge the reality of the implementing abstractions.

The obvious place to look for sciences building upon other sciences is the hierarchy of the sciences. To take the most concrete case, chemistry is built on the abstraction of the atom as an entity with an internal structure. Yet we certainly don't wonder about whether chemistry is a real science.

Molecules are the (emergent) entities of chemistry. They form when combinations of atoms in are in a lower energy state than the atoms would be in isolation. How does nature implement this? It does it in terms of abstract structures known as orbitals.

Molecules form when orbitals from pairs of atoms merge. What is an orbital? It is part of the abstract design—the design that determines how electrons and protons relate to each other—that matter implements by following the rules of quantum mechanics. Thus molecular bonds are implemented by nature through the quantum mechanical mechanisms of orbitals. Like the formation of atoms themselves, chemical bonding is part of the free lunch that nature sets out for us—and another illustration that emergence is a fundamental aspect of nature.

Of course, this is just one example. As we shall saw above, the complexity that we see around us is a direct result of the fact that new abstractions may be built on top of the functionalities provided by existing abstractions.

# 8  Varieties of Emergence

In this section we stand back and review the kinds of emergence we have discussed. In particular we discuss two categories of emergence: *static emergence* and *dynamic emergence*. We also suggest that these categories correspond to Weinberg's notion of petty and grand reductionism.

As in the case of Weinberg's petty and grand reductionism, static emergence, while of great importance, is of lesser interest. It is dynamic emergence, and especially stigmergic dynamic emergence that is central to complex systems.

Recall that we defined a phenomenon as emergent over a underlying model if (a) it has an independent conceptualization and (b) it can be implemented in terms of elements of that model.

## 8.1  Static emergence

An emergent phenomenon is statically emergent if its implementation does not depend on time.

As an interesting example of static emergence, consider cloth as a collection of threads woven together. Cloth has the emergent property that it is able to cover a surface. This property is implicitly two dimensional. The components of cloth, i.e., threads, do not have (or at least are not understood in terms of) that property. A thread is understood in terms of the property length. Yet when threads are woven together the resulting cloth has this new property, which effectively converts a collection of one dimensional components to a two dimensional object.

Many human manufactured or constructed artifacts exhibit static emergence. A house has the statically emergent property number-of-bedrooms. More generally, a house has the emergent property that it can serve as a resi-dence. Static emergence also occurs in nature. As Weinberg points out,

> [A] diamond is hard because the carbon atoms of which it is composed can fit together neatly [even though] it doesn't make sense to talk about the hardness ... of individual 'elementary' particles.

Since statically emergent phenomena must be implemented in terms of some underlying model, and since time is by definition excluded from that implementation, static emergence is equivalent to Weinberg's petty reductionism.

## 8.2  Dynamic emergence

Properties or phenomena of a model are dynamically emergent if they are defined in terms of how the model changes (or doesn't change) over some time. Dynamic emergence occurs either with or without autonomous entities. We call the former stigmergic emergence, but we look at non-stigmergic dynamic emergence first.

## 8.3  Non-stigmergic dynamic emergence

Interactions among at-equilibrium entities result in non-stigmergic dynamic emergence. Two examples are: (a) objects moving in space and interacting according to Newtonian mechanics and (b) the quantum wave function.

What appears to be distinctive about such systems is that they are not characterized in terms of discrete states that their elements assume. Elements do not transition from one state to another. Such systems may be defined in terms of continuous equations.

The quantum wave function is an especially interesting example. As long as it does not undergo decoherence, i.e., interaction with an environment, the wave function encompasses all possibilities, but it realizes none of them.

Since quantum states are discrete (hence the term *quantum*), objects cannot transition smoothly from one quantum state to another. So how does that transition occur? Quantum theory turns these transitions into probabilities. As Hardy points out [Hardy], by making such transitions probabilistically continuous,

> quantum theory offers us a way to have the advantages of discreteness and continuity at the same time.

For all practical purposes, actually to assume a state requires something more, the so-called collapse of the wave function. That happens stigmergically. At the quantum level, stigmergy is equivalent to decoherence.

## 8.4  Dynamic emergence and grand reductionism

As static emergence corresponds to Weinberg's petty reductionism, dynamic emergence seems to correspond nicely to Weinberg's grand reductionism. Weinberg explains grand reductionism as follows

> [T]he reductionist regards the general theories governing air and water and radiation as being at a deeper level than theories about cold fronts or thunderstorms, not in the sense that they are more useful, but only in the sense that the latter can in principle be understood as mathematical consequences of the former. The reductionist program of physics is the search for the common source of all explanations. ...

We hope that Weinberg would not object to the following paraphrase.

> The reductionist goal (with respect to reducing weather terminology to the terminology of physics) is to build a model (a) whose elements include air, water vapor, and radiation and (b) whose elements interact according to the principles of physics. When that model is run it will generate the emergent phenomena that we would recognize as cold fronts and thunderstorms.

Grand reductionism is thus the explanation of phenomena at one level in terms of phenomena at a more fundamental level. One shows that when the laws at the more fundamental level are applied, the result will be the phenomena of interest at the less fundamental level. Since these sorts of models are inevitably dynamic this is dynamic emergence but expressed in other terms.

Nor should the equating of grand reductionism with dynamic emergence surprise anyone in the field of complex systems. After all, the presumed reason to build a model is to show that a set of lower level rules will produce higher level results—which is exactly the grand reductionist agenda.

Of course the terminology that we would use is not just that lower level rules produce higher level results, that lower level rules may implement a higher level abstraction.

But no matter how it is expressed, without the lower level substrate, the higher level phenomena would not exist.

## 8.5  Stigmergic emergence

Stigmergic emergence is dynamic emergence that involves autonomous entities. What tends to be most interesting about autonomous entities are (a) they may assume discrete states and (b) they change state as they interact with their environments.[44]

Furthermore, not only do autonomous entities depend on their environments as sources of energy and other resources, the environment on which any autonomous entity depends includes other autonomous entities. Of course these other autonomous entities also depend on their environment, etc. These dependencies form networks of enormous and complexity in which the dependency links are frequently not higher depending on lower.

Static and non-stigmergic dynamic emergence is fairly well-behaved. One can often write down equations that characterize entire systems in which it occurs—even though it may not be practical to solve those equations for other than trivial cases. Stigmergic emergence is far worse. Because of the relative interdependence of the components, it is virtually impossible to provide a global equation-like characterization of the system as a whole. Stigmergic emergence is the source of the complexity in nature. It is because of stigmergic emergence that complex systems are complex.

This would seem to put a final stake in the heart of Laplace's demon, the hypothetical computing device that if given details about the initial state of the universe would be able to compute all future states. Laplace's demon may succeed in a Newtonian universe, for which it was invented. Laplace's demon may even succeed in a quantum mechanical universe in that the quantum wave equation is deterministic—even though it characterizes probability amplitudes and hence its collapse is not. But if nature includes asynchronously acting autonomous entities, some of which may them-

---

[44] One might liken an isolated quantum wave system to the inside of an autonomous entity. It assumes a

state (i.e., collapses) when it interacts with its environment.

selves embody quantum probability transitions, many of which are mutually interdependent, and all of which depend on their environment, which includes other autonomous entities for their operation and persistence, Laplace's demon will be way beyond its depth.

One possible simple formal model for such a computational system is a shared tape Turing Machine community: a collection of asynchronously operating Turing Machines that share a single tape.[45]

Some proponents of agent-based modeling argue for that approach on the grounds that even though some domains may have global characterizations, those characterizations are much too complex to compute. Our position is that agent-based modeling is appropriate because that's how nature is.

# 9   Some practical considerations

## 9.1   *Emergence and software*
As noted earlier, the computation that results when software is executed is emergent. It is an epiphenomenon of the operation of the (actual or virtual) machine that executes the software.

Earlier we defined emergence as synonymous with epiphenomenon. At that time we suggested that formalizable epiphenomena are often of significant interest. We also said that formalization may not always be in the cards. Software, which one would imagine to be a perfect candidate for formalization, now seems to be a good example of an epiphenomenon that is unlikely to be formalized.

It had once been hoped that software development could evolve to a point at which one need only write down a formal specification of what one wanted the software to do. Then some automatic process would produce software that satisfied that specification.

That dream now seems quite remote. Besides the difficulty of developing (a) a satisfactory specification language and (b) a system that can translate specifications written in such a language into executable code, the real problem is that it has turned out to be at least as difficult and complex to write formal specifications as it is to write the code that produces the specified results.

Even if one could write software by writing specifications, in many cases—especially cases that involve large and complex systems, the kinds of cases for which it really matters—doing so doesn't seem to result in much intellectual leverage, if indeed it produces any at all.

This illustrates quite nicely that we often find ourselves in the position of wanting to produce epiphenomena (epiphenomena, which may be very important to us), whose formalization as an abstraction we find to be either infeasible or not particularly useful.

## 9.2   *Bricolage as design*
The process of building one capability on top of another not only drives the overall increase in complexity, it also provides guidance to designers about how to do good design work. Any good designer—a developer, an architect, a programmer, or an engineer—knows that it is often best if one can take advan-tage of forces and processes already in existence as part of one's design.

---

[45]   Wegner's work [Wegner] on non-traditional Turing Machine models begins to explore his own models. Cockshott and Michaelson [Cockshott] dispute whether Wegner's models extend the power of the Turing machine.

But even before engineering, we as human beings made use of pre-existing capabilities. Agriculture and animal husbandry use both plant reproduction and such animal capabilities as locomotion or material (i.e., skin) production for our own purposes. The exploitation of existing capabilities for our own purposes is not a new idea.

An interesting example of this approach to engineering involves recent developments in robotics. Collins reported [Collins] that a good way to make a robot walk is by exploiting gravity through what he called passive-dynamic motion—raise the robot's leg and let gravity pull it back down—rather than by directing the robot's limbs to follow a predefined trajectory.

This illustrates in a very concrete way the use of an existing force in a design. Instead of building a robot whose every motion was explicitly programmed, Collins built a robot whose motions were controlled in part by gravity, a pre-existing force.

### 9.3 Infrastructure-centric development

Building new capabilities on top of existing ones is not only good design, it is highly leveraged design. But now that we are aware of this strategy a further lesson can be drawn. New systems should be explicitly designed to serve as a possible basis for systems yet to come. Another way of putting this is that every time we build a new system, it should be built so that it becomes part of our environment, i.e., our infrastructure, and not just a piece of closed and isolated functionality.

By infrastructure we mean systems such as the Internet, the telephone system, the electric power distribution system, etc. Each of these systems can be characterized in isolation in terms of the particular functions they perform. But more important than the functional characterization of any of these individual systems is the fact that they exist in the environment in such a way that other systems can use them as services.

We should apply this perspective to all new systems that we design: design them as infrastructure services and not just as bits of functionality. Clearly Microsoft understands this. Not only does it position the systems it sells as infrastructure services, it also maintains tight ownership and control over them. When such systems become widely used elements of the economy, the company makes a lot of money. The tight control it maintains and the selfishness with which it controls these systems earns it lots of resentment as well. Society can't prosper when any important element of its infrastructure is controlled primarily for selfish purposes.

The US Department of Defense (DoD) is currently reinventing itself [Dick] to be more infrastructure-centric. This requires it to transform what is now a

> huge collection of independent "stovepipe" information systems, each supporting only its original procurement specification, to a unified assembly of interoperating systems.

The evocative term *stovepipe* is intended to distinguish the existing situation—in which the DoD finds that it has acquired and deployed a large number of functionally isolated systems (the "stovepipes")—from the more desirable situation in which all DoD systems are available to each other as an infrastructure of services.

## 9.4 Service refactoring and the age of services

The process whereby infrastructure services build on other infrastructure services leads not only to new services, it also leads to service refactoring. The corporate trend toward outsourcing functions that are not considered part of the core competence of the corporation illustrates this. Payroll processing is a typical example.

Because many organizations have employees who must be paid, these organizations must provide a payroll service for themselves. It has now become feasible to factor out that service and offer it as part of our economic infrastructure.

This outsourcing of internal processes leads to economic efficiencies in that many such processes can be done more efficiently when performed by specialized organizations. Such specialized organizations can take advantage of economies of scale. They can also serve as focal points where expertise in their specialized service can be concentrated and the means of providing those services improved.

As this process establishes itself ever more firmly, more and more organizations will focus more on offering services rather than functions, and organizations will become less stovepiped.

We frequently speak of the "service industries." For the most part this term has been used to refer to low level services—although even the fast food industry can be seen as the "outsourcing" of the personal food preparation function. With our more general notion of service in mind, historians may look back to this period as the beginning of the age of services.

Recall that a successful service is an autonomous entity. It persists as long as it is able to extract from its environment enough resources, typically money, to perpetuate itself.

## 9.5 A possible undesirable unintended consequence

The sort of service refactoring we just discussed tends to make the overall economic system more efficient. It also tends to improve reliability: the payroll service organizations are more reliable than the average corporate payroll department.

On the other hand, by eliminating redundancy, efficiency makes the overall economic system more vulnerable to large scale failure. If a payroll service organization has a failure, it is likely to have a larger impact than the failure of any one corporate payroll department. This phenomenon seems to be quite common—tending to transform failure statistics from a Gaussian to a scale free distribution: the tails are longer and fatter. [Colbaugh]  Failures may be less frequent, but when they occur they may be more global.

This may be yet another unintended and unexpected emergent phenomenon—a modern example of the tragedy of the commons. Increased economic efficiency leads to increased vulnerability to major disasters at the societal-level.

On the other hand, perhaps our growing realization that catastrophic failures may occur along with our ability to factor out commonly needed services will help us solve this problem as well. We now see increasing number of disaster planning services being offered.

## 9.6 Modeling: the difficulty of looking downward

The perspective we have described yields two major implications for modeling. We refer to them as the *difficulty of looking downwards* and *the difficulty of looking upwards*. In both cases, the problem is that it is very difficult to model significant creativity—notwithstanding the fact that surprises do appear in some of our models. In this section we examine the difficulty of looking downward. In the next we consider the difficulty of looking upward.

Strict reductionism, our conclusion that all forces and actions are epiphenomenal over forces and actions at the fundamental level of physics, implies that it is impossible to find a non-arbitrary base level for models. One never knows what unexpected effects one may be leaving out by defining a model in which interactions occur at some non-fundamental level.

Consider a model of computer security. Suppose that by analyzing the model one could guarantee that a communication line uses essentially unbreakable encryption technology. Still it is possible for someone inside to transmit information to someone outside.

How? By sending messages in which the content of the message is ignored but the frequency of transmission carries the information, e.g., by using Morse code. The problem is that the model didn't include that level of detail. This is the problem of looking downward.

A further illustration of this difficulty is that there are no good models of biological arms races. (There don't seem to be any good models of significant co-evolution at all.) There certainly are models of population size effects in predator-prey simulations. But by bio-logical arms races we are talking about not just population sizes but actual evolutionary changes.

Imagine a situation in which a plant species comes under attack from an insect species. In natural evolution the plant may "figure out" how to grow bark. Can we build a computer model in which this solution would emerge? It is very unlikely. To do so would require that our model have built into it enough information about plant biochemistry to enable it to find a way to modify that biochemistry to produce bark, which itself is defined implicitly in terms of a surface that the insect cannot penetrate. Evolving bark would require an enormous amount of information—especially if we don't want to prejudice the solution the plant comes up with.

The next step, of course, is for the insect to figure out how to bore through bark. Can our model come up with something like that? Unlikely. What about the plant's next step: "figuring out" how to produce a compound that is toxic to the insect? That requires that the model include information about both plant and insect biochemistry—and how the plant can produce a compound that interferes with the insect's internal processes. This would be followed by the development by the insect of an anti-toxin defense.

To simulate this sort of evolutionary process would require an enormous amount of low level detail—again especially if we don't want to prejudice the solution in advance.

Other than Tierra (see [Ray]) and its successors, which seem to lack the richness to get very far off the ground, as far as we know, there are no good computer models of biological arms races. A seemingly promising approach would be an agent-based system in which each agent ran its own internal genetic programming model. But we are unaware of any such work.[46]

Finally, consider the fact that geckos climb walls by taking advantage of the Van der Walls "force." (We put *force* in quotation marks because there is no Van der Walls force. It is an epiphenomenon of relatively rarely occurring quantum phenomena.) To build a model of evolution in which creatures evolve to use the Van der Walls force to climb walls would require that we build quantum physics into what is presumably intended to be a relatively high-level biological model in which macro geckos climb macro walls

It's worth noting that the use of the Van der Walls force was apparently not an extension of some other gecko process. Yet the gecko somehow found a way to reach directly down to a quantum-level effect to find a way to climb walls.

The moral is that any base level that we select for our models will be arbitrary, and by choosing that base level, we may miss important possibilities. Another moral is that models used when doing computer security or terrorism analy-

sis—or virtually anything else that includes the possibility of creative adaptation—will always be incomplete. We will only be able to model effects on the levels for which our models are defined. The imaginations of any agents that we model will be limited to the capabilities built into the model.

### 9.7 Modeling: the difficulty of looking upward

We noted earlier that when a glider appears in the Game of Life, it has no effect on the how the system behaves. The agents don't see a glider coming and duck. More significantly we don't know how to build systems so that agents will be able to notice gliders and duck.

It would be an extraordinary achievement in artificial intelligence to build a modeling system that could notice emergent phenomena and see how they could be exploited. Yet we as human beings do this all the time. The dynamism of a free-market economy depends on our ability to notice newly emergent patterns and to find ways to exploit them.

Al Qaeda noticed that our commercial airlines system can be seen as a network of flying bombs. Yet no model of terrorism that doesn't have something like that built into it will be able to make that sort of creative leap. Our models are blind to emergence even as it occurs within them.

Notice that this is not the same as the difficulty of looking downward. In the Al Qaeda example one may assume that one's model of the airline system includes the information that an airplane when loaded with fuel will explode when it crashes. The creative leap is to notice that one can use that phenomenon for new purposes. This is easier than the

---

[46] Genetic programming is relevant because we are assuming that the agent has an arbitrarily detailed description of how the it functions and how elements in its environment function.

Notice how difficult it would be implement such a system. The agent's internal model of the environment would have to be updated continually as the environment changed. That requires a means to perceive the environment and to model changes in it. Clearly that's extraordinarily sophisticated. Although one could describe such a system without recourse to the word *consciousness*, the term does come to mind.

Nature's approach is much simpler: change during reproduction and see what happens. If the result is unsuccessful, it dies out; if it is successful it persists and reproduces. Of course that requires an entire generation for each new idea.

problem of looking downward. But it is still a very difficult problem.

The moral is the same as before. Models will always be incomplete. We will only be able to model effects on the levels for which our models are defined. The imaginations of any agents that we model will be limited to the capabilities built into the model.

# 10 Concluding remarks

## 10.1 *Computer Science and Philosophy*

- For centuries after Newton, nature was seen as a perfect clockwork mechanism.

- After the invention of the telephone, the brain was likened to a switchboard.

Science and technology tends to shape our view of the world. It's not surprising that the perspectives developed in this article reflect those of Computer Science, the predominant perspective of this age. Is this parochialism? It's difficult to tell from so close. One thing is clear. Because Computer Science has wrestled—with some success—with many serious intellectual challenges, it is not unreasonable to hope that the field may contribute something to the broader intellectual community.

It is useful to think of computers as reification machines: they make the abstract concrete. As such they are similar to the drawing tools of an animator—who draws anything that crosses his or her mind, no matter how wild or fanciful. In the hands of a skilled and creative programmer,[47] a computer and an associated software development environment invite virtually unlimited creativity.

But there is a major difference between the product produced by an animator and that produced by a programmer. An animator produces images that have meaning only when they enter the mind of another human being. They do not stand on their own. They are meant strictly as communication from one human being to another. Nor are either the physical medium on which the images are stored or the mechanisms that causes the images to be displayed of much significance. The only thing that matters is that the images arrive as intended in the minds of the viewers.

A computer program is stuck in the real world. It's work is to shape the activity of a computer, a physical device. Many programs execute without even being observed by people—programs in cars and other machinery, for example. And even programs that perform strictly symbolic transformations, do work that may be understood as constraining the forces of nature (the motion of electrons) to some formal shape.

To make the computer a useful reification device—to make it possible for programmers to write text that causes a computer to convert programmers' fantasies to some concrete form—Computer Science has had to deal with some of philosophy's thorniest issues.

- Computer Science has created languages that are both formally de-

---

[47] We use the term *programmer* deliberately. Fancier terms like *software developer* (or *software engineer* or worse *information technology specialist*) lose the focus of what is really happening when one creates software. A programmer (writes text

that) programs, i.e., shapes, the way a computer behaves. That's really all that matters: what does the text tell the computer to do?

In the preceding, note the stigmergy and downward entailment. A program doesn't *tell* a computer anything. Successful programmers work from that perspective all the time.

fined—with formal syntax and semantics—and operational, i.e., they actually function in the real world.

- Computer Science has figured out how to represent information in databases in ways that allow that information to hang together meaningfully.

- Computer Science has faced—and to a significant extent resolved—the problem of working on many levels of abstraction and complexity simultaneously.

If insights gained from these and other intellectual wrestling matches can be applied in a wider context, it is only Computer Science paying back the debt that it owes to the engineers, scientists, mathematicians, and philosophers who set the stage for and participated in its development.

## 10.2 Constructive science

For most of its history, science has pursued the goal of explaining existing phenomena in terms of simpler phenomena. That's the reductionist agenda.

The approach we have taken is to ask how new phenomena may be constructed from and implemented in terms of existing phenomena. That's the creative impulse of artists, computer scientists, engineers—and of nature. It is these new phenomena that are often thought of as emergent.

When thinking in the constructive direction, a question arises that is often underappreciated: what allows one to put existing things together to get something new—and something new that will persist in the world? What binding forces and binding strategies do we (and nature) have at our disposal?

Our answer has been that there are two sorts of binding strategies: energy wells and energy-consuming processes. Energy wells are reasonably well understood—although it is astonishing how many different epiphenomena nature and technology have produced through the use of energy wells.

We have not even begun to catalog the ways in which energy-consuming processes may be used to construct stable, self-perpetuating, autonomous entities.

Earlier we wrote that science does not consider it within its realm to ask constructivist questions. That is not completely true. Science asks about how we got here from the big bang, and science asks about biological evolution. These are both constructivist questions. Since science is an attempt to understand nature, and since constructive processes occur in nature, it is quite consistent with the overall goals of science to ask how these constructive processes work. As far as we can determine, there is no subdiscipline of science that asks, in general, how the new arise from the existing.

Science has produced some specialized answers to this question. The biological evolutionary explanation involves random mutation and crossover of design records. The cosmological explanation involves falling into energy wells of various sorts. Is there any more to say about how nature finds and then explores new possibilities? If as Dennett argues in [Dennett '96] this process may be fully explicated as generalized Darwinian evolution, questions still remain. Is there any useful way to characterize the search space that nature is exploring? What search strategies does nature use to explore that space? Clearly one strategy is human inventiveness.

# 11 Acknowledgement

We are grateful for numerous enjoyable and insightful discussions with Debora Shuger during which many of the ideas in this paper were developed and refined.

We also wish to acknowledge the following websites and services, which we used repeatedly.

- Google (www.google.com);
- The Stanford Encyclopedia of Philosophy (plato.stanford.edu);
- OneLook Dictionary Search (onelook.com).

# 12 Appendix. Game of Life Patterns

Intuitively, a Game of Life pattern is the step-by-step time and space progression on a grid of a discernable collection of inter-related live cells. We formalize that notion in three steps.

1. First we define a static construct called the live cell group. This will be a group of functionally isolated but internally interconnected cells.

2. Then we define Game of Life *basic patterns* as temporal sequences of live cell groups. The Game of Life *glider* and *still-life* patterns are examples

3. Finally we extend the set of patterns to include combinations of basic patterns. The more sophisticated Game of Life patterns, such the *glider gun*, are examples.

## 12.1 Live cell groups

The fundamental construct upon which we will build the notion of a pattern is what we shall call a *live cell group*.

A live cell group is a collection of live and dead cells that have two properties.

1. They are functionally isolated from other live cells.

2. They are functionally related to each other.

More formally, we define cells $c_0$ and $c_n$ in a Game of Life grid to be *connected* if there are cells $c_1$, $c_2$, …, $c_{n-1}$ such that for all $i$ in $0 .. n-1$

1. $c_i$ and $c_{i+1}$ are *neighbors*, as defined by Game of Life, and

2. either $c_i$ or $c_{i+1}$ (or both) are *alive*, as defined by Game of Life.

Connectedness is clearly an equivalence relation (reflexive, symmetric, and transitive), which partitions a Game of Life board into equivalence classes of cells. Every dead cell that is not adjacent to a live cell (does not have a live cell as a Game of Life neighbor) becomes a singleton class.

Consider only those connectedness equivalence classes that include at least one live cell. Call such an equivalence class a *live cell group* or *LCG*.

Define the *state* of an LCG as the specific configuration of live and dead cells in it. Thus, each LCG has a state.

No limitation is placed on the size of an LCG. Therefore, if one does not limit the size of the Game of Life grid, the number of LCGs is unbounded.

Intuitively, an LCG is a functionally isolated group of live and dead cells, contained within a boundary of dead cells. Each cell in an LCG is a neighbor to at least one live cell within that LCG.

As a consequence of this definition, each live cell group consists of an "inside," which contains all its live cells (possibly along with some dead cells), plus a "surface" or "boundary" of dead cells. (The surface or boundary is also considered part of the LCG.)

## 12.2 Basic patterns: temporal sequences of live cell groups

Given this definition, we can now build temporal sequences of LCGs. These will be the Game of Life basic patterns.

The Game of Life rules define transitions for the cells in a LCG. Since an LCG is functionally isolated from other live cells, the new states of the cells in an LCG are determined only by other cells in the same LCG.[48]

Suppose that an LCG contains the only live cells on a Game of Life grid. Consider what the mapping of that LCG by the Game of Life rules will produce. There are three possibilities.

1. The live cells may all die.

2. The successor live cells may consist of a single LCG—as in a *glider* or *still life*.

3. The successor live cells may partition into multiple LCGs—as in the so-called *bhepto* pattern, which starts as a single LCG and eventually stabilizes as 4 *still life* LCGs and two *glider* LCGs.

In other words, the live cells generated when the Game of Life rules are applied to an LCG will consist of 0, 1, or multiple successor LCGs.

More formally, if $\ell$ is an LCG, let Game of Life($\ell$) be the set of LCGs that are formed by applying the Game of Life rules to the cells in $\ell$. For any particular $\ell$, Game of Life($\ell$) may be empty; it may be contain a single element; or it may contain multiple elements. If $\ell'$ is a member of Game of Life($\ell$) write $\ell \rightarrow \ell'$.

For any LCG $\ell_0$, consider a sequence of successor LCGs generated in this manner:

$$\ell_0 \rightarrow \ell_1 \rightarrow \ell_2 \rightarrow \ell_3 \rightarrow \dots \, .$$

Extend such a sequence until one of three conditions occurs.

1. There are no successor LCGs, i.e., Game of Life($\ell_i$) is empty—all the live cells in the final LCG die. Call these terminating sequences.

2. There is a single successor LCG, i.e., Game of Life($\ell_i$) = {$\ell_k$}, but that successor LCG is in the same state as an LCG earlier in the sequence, i.e., $\ell_k = \ell_j$, $j < k$. Call these repeating sequences.

3. The set Game of Life($\ell_i$) of successor LCGs contains more than one LCG, i.e., the LCG branches into two or more LCGs. Call these branching sequences.

Note that some LCG sequences may never terminate. They may simply produce larger and larger LCGs. The so-called *spacefiller* pattern, which actually consists of multiple interacting LCGs, one of which fills the entire grid with a single LCG as it expands,[49] is an amazing example of such a pattern. I do not know if there is an LCG that expands without limit on its own. If any such exist, call these infinite sequences.

For any LCG $\ell_0$, if the sequence

---

[48] In particular, no LCG cells have live neighbors that are outside the LCG. Thus no cells outside the LCG need be considered when determining the GoL transitions of the cells in an LCG. A dead boundary cell may become live at the next time-step, but it will do so only if three of its neighbors within the LCG are live. Its neighbors outside the LCG are guaranteed to be dead.

If a boundary cell does become live, the next-state LCG of which it is a member will include cells that were not part of its predecessor LCG.

---

[49] See the spacefiller pattern on http://www.math.com/students/wonders/life/life.html or http://www.ibiblio.org/lifepatterns.

$$\ell_0 \to \ell_1 \to \ell_2 \to \ell_3 \to \dots .$$

is finite, terminating in one of the three ways described above, let seq($\ell_0$) be that sequence along with a description of how it terminates. If

$$\ell_0 \to \ell_1 \to \ell_2 \to \ell_3 \to \dots .$$

is infinite, then seq($\ell_0$) is undefined.

Let BP (for Basic Patterns) be the set of finite non-branching sequences as defined above. That is,

$$BP = \{ seq(\ell_0) \mid \ell_0 \text{ is an LCG} \}$$

Note that it is not necessary to extend these sequences backwards. For any LCG $\ell_0$, one could define the pre-image of $\ell_0$ under the Game of Life rules. Game of Life$^{-1}(\ell)$ is the set of LCGs $\ell'$ such that Game of Life($\ell'$) = $\ell$.

For any chain seq($\ell_0$) in BP, one could add all the chains constructed by prefixing to seq($\ell_0$) each of the predecessors $\ell'$ of $\ell_0$ $\ell'$ as long as $\ell'$ does not appear in seq($\ell'$). But augmenting BP in this way would add nothing to BP since by definition seq($\ell'$) is already defined to be in BP for each $\ell'$.

We noted above that we do not know if there are unboundedly long sequences of LCGs beginning with a particular $\ell_0$. With respect to unboundedly long predecessor chains, it is known that such unbounded predecessor chains (of unboundedly large LCGs) exist. The so-called *fuse* and *wick* patterns are LCG sequences that can be extended arbitrarily far backwards.[50] When run forward such fuse or wick LCGs converge to a single LCG. Yet given the original definition of BP even these LCG sequences are included in it. Each of these unbounded predecessor chains is included in BP starting at each predecessor LCG.

Clearly BP as defined includes many redundant pattern descriptions. No attempt is made to minimize BP either for symmetries or for overlapping patterns in which one pattern is a suffix of another—as in the fuse patterns. In a computer program that generated BP, such efficiencies would be important.

### 12.3 BP is recursively enumerable

The set BP of basic Game of Life patterns may be constructed through a formal iterative process. The technique employed is that used for the construction of many recursively enumerable sets.

1. Generate the LCGs in sequence.

2. As each new LCG is generated, generate the next step in each of the sequences starting at each of the LCGs generated so far.

3. Whenever an LCG sequence terminates according to the BP criteria, add it to BP.

The process sketched above will effectively generate all members of BP. Although theoretically possible, such a procedure will be so inefficient that it is useless for any practical purpose.[51] The die; the remaining cells remain alive. A simple fuse pattern may be augmented by adding more complex features at one end, thereby building a pattern that becomes active when the fuse exhausts itself. Such a pattern can be built with an arbitrarily long fuse.

---

[50] A simple fuse pattern is a diagonal configuration of live cells. At each time step, the two end cells

[51] Many much more practical and efficient programs have been written to search for patterns in the GoL and related cellular automata. See http://www.ics.uci.edu/~eppstein/ca/search.html for a list of such programs.

only reason to mention it here is to establish that BP is recursively enumerable. Whether BP is recursive depends on whether one can in general establish for any LCG $\ell_0$ whether seq($\ell_0$) will terminate.

### 12.4 Game of Life patterns: combinations of basic patterns

Many of the interesting Game of Life patterns arise from interactions between and among basic patterns. For example, the first pattern that generated an unlimited number of live cells, the glider gun, is a series of interactions among combinations of multiple basic patterns that cyclically generate gliders.

To characterize these more complex patterns it is necessary to keep track of how basic patterns interact. In particular, for each element in BP, augment its description with information describing

a) its velocity (rate, possibly zero, and direction) across the grid,

b) if it cycles, how it repeats, i.e., which states comprise its cycle, and

c) if it branches, what the offspring elements are and where they appear relative to final position of the terminating sequence.

Two or more distinct members of BP that at time step i are moving relative to each other may interact to produce one or more members of BP at time step i+1.

The result of such a BP "collision" will generally depend on the relative positions of the interacting basic patterns. Even though the set BP of basic patterns is infinite, since each LCG is finite, by using a technique similar to that used for generating BP itself, one can (very tediously) enumerate all the possible BP interactions.

More formally, let $\mathscr{P}_f$(BP) be the set of all finite subsets of BP. For each member of $\mathscr{P}_f$(BP) consider all possible (still only a finite number) relative configurations of its members on the grid so that there will be some interaction among them at the next time step. One can then record all the possible interactions among finite subsets of BP.

These interactions would be equivalent to the APIs for the basic patterns. We could call a listing of them BP-API. Since BP is itself infinite, BP-API would also be infinite. But BP-API would be effectively searchable. Given a set of elements in BP, one could retrieve all the interactions among those elements. BP-API would then provide a documented starting point for using the Game of Life as a programming language.

As in traditional programming languages, as more complex interactions are developed, they too could be documented and made public for others to use.

# Figures and Tables

**Table 1. Dissipative structures vs. self-perpetuating entities**

| Dissipative structures | Self-perpetuating entities |
|---|---|
| Pure epiphenomena, e.g., 2-chamber example. | Has functional design, e.g., hurricane. |
| Artificial boundaries. | Self-defining boundaries |
| Externally maintained energy gradient. | Imports, stores, and internally distributes energy. |

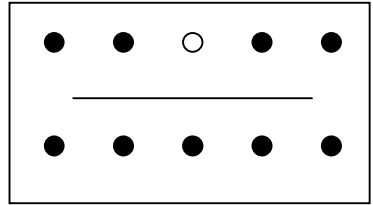

**Figure 1.** Bit 3 off  and then on.

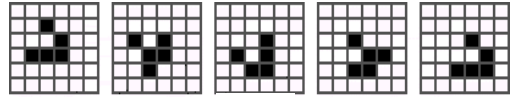

**Figure 2.** A glider

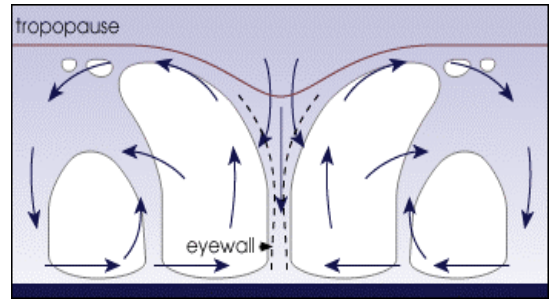

**Figure 3.** Anatomy of a hurricane. [Image from [NASA].]